\documentclass[aps,prapplied,reprint,superscriptaddress,longbibliography]{revtex4-2}

\usepackage{graphicx}
\usepackage{color,amssymb,amsmath}
\usepackage[normalem]{ulem}
\usepackage{hyperref}
\usepackage{lipsum}
\usepackage{physics}
\usepackage{float}

\makeatletter
\newcommand*{\rom}[1]{\expandafter\@slowromancap\romannumeral #1@}
\makeatother

\begin{document}
\title{Flip-chip-based fast inductive parity readout of a planar superconducting island}

\author{M.~Hinderling}
\affiliation{IBM Research Europe - Zurich, S\"aumerstrasse 4, 8803 R\"uschlikon, Switzerland}
	
\author{S.~C.~ten~Kate}
\affiliation{IBM Research Europe - Zurich, S\"aumerstrasse 4, 8803 R\"uschlikon, Switzerland}

\author{D.~Z.~Haxell}
\affiliation{IBM Research Europe - Zurich, S\"aumerstrasse 4, 8803 R\"uschlikon, Switzerland}
	
\author{M.~Coraiola}
\affiliation{IBM Research Europe - Zurich, S\"aumerstrasse 4, 8803 R\"uschlikon, Switzerland}

\author{S.~Paredes}
\affiliation{IBM Research Europe - Zurich, S\"aumerstrasse 4, 8803 R\"uschlikon, Switzerland}

\author{E.~Cheah}
\affiliation{Solid State Physics Laboratory, ETH Zurich, Otto-Stern-Weg 1, 8093 Z\"urich, Switzerland}

\author{F. Krizek}
\altaffiliation[Present address: ]{Institute of Physics, Czech Academy of Sciences, 162 00 Prague, Czech Republic}
\affiliation{IBM Research Europe - Zurich, S\"aumerstrasse 4, 8803 R\"uschlikon, Switzerland}
\affiliation{Solid State Physics Laboratory, ETH Zurich, Otto-Stern-Weg 1, 8093 Z\"urich, Switzerland}

\author{R. Schott}
\affiliation{Solid State Physics Laboratory, ETH Zurich, Otto-Stern-Weg 1, 8093 Z\"urich, Switzerland}

\author{W.~Wegscheider}
\affiliation{Solid State Physics Laboratory, ETH Zurich, Otto-Stern-Weg 1, 8093 Z\"urich, Switzerland}

\author{D.~Sabonis}
\affiliation{IBM Research Europe - Zurich, S\"aumerstrasse 4, 8803 R\"uschlikon, Switzerland}

\author{F.~Nichele}
\email{Author to whom correspondence should be addressed: fni@zurich.ibm.com}
\affiliation{IBM Research Europe - Zurich, S\"aumerstrasse 4, 8803 R\"uschlikon, Switzerland}

\date{\today}

\begin{abstract}
Properties of superconducting devices depend sensitively on the parity (even or odd) of the quasiparticles they contain. Encoding quantum information in the parity degree of freedom is central in several emerging solid-state qubit architectures. Yet, accurate, non-destructive, and time-resolved parity measurement is a challenging and long-standing issue.
Here we report on control and real-time parity measurement in a superconducting island embedded in a superconducting loop and realized in a hybrid two-dimensional heterostructure using a microwave resonator. Device and readout resonator are located on separate chips, connected via flip-chip bonding, and couple inductively through vacuum. The superconducting resonator detects the parity-dependent circuit inductance, allowing for fast and non-destructive parity readout. We resolved even and odd parity states with signal-to-noise ratio SNR $\approx3$ with an integration time of 20~$\mu$s and detection fidelity exceeding 98~$\%$. Real-time parity measurement showed state lifetime extending into millisecond range. Our approach will lead to better understanding of coherence-limiting mechanisms in superconducting quantum hardware and provide novel readout schemes for hybrid qubits. 
\end{abstract}

\keywords{parity, flip-chip, hybrid, superconductivity, inductive, readout}
\maketitle

\section*{Introduction}

Control and measurement of isolated quantum systems is a crucial task for the realization of practical quantum devices~\cite{RevModPhys.93.025005, PRXQuantum.2.040202}. Within the last decades, quantum dots emerged as an ideal platform for the realization of critical components for charge~\cite{kim2015microwave,petersson2010quantum} and spin~\cite{weinstein2023universal,philips2022universal} qubits due to their ability to trap, control and sense individual charge carriers using electrostatic means. Superconducting quantum dots posses an additional degree of freedom, given by the parity (even or odd) of unpaired excitations~\cite{tuominen1992experimental,aumentado2004nonequilibrium,PhysRevB.57.120, naaman2006time,PhysRevLett.97.106603, van2006supercurrent}. Parity is an attractive alternative to charge and spin for encoding long-lived quantum states, for example in $0-\pi$~\cite{PhysRevA.87.052306}, topological~\cite{aasen2016milestones}, and Kerr cat qubits~\cite{grimm2020stabilization}. Performing real-time, non-destructive charge parity readout is therefore a fundamental requirement in proposed topological quantum computing schemes~\cite{PhysRevB.94.155417,dai2015extracting,karzig2017scalable,plugge2017majorana,szechenyi2020parity,PhysRevB.107.L121401} and Andreev spin qubits~\cite{PhysRevX.9.011010,hays2021coherent,pita2023direct}. More generally, continuous monitoring of the parity of superconducting systems could shine new light on parity dynamics and on fundamental mechanisms limiting coherence in superconducting hardware.

Investigating effects related to parity in superconductor-semiconductor hybrid systems is a subject of intense research \cite{wang2022supercurrent,PhysRevB.100.020502, nguyen2022electrostatic, higginbotham2015parity}. However, direct measurements of the parity in superconducting systems is often challenging. Successful demonstrations made use of parity-dependent switching supercurrents~\cite{eiles1993even, van2006supercurrent, van2015one,szombati2016josephson, razmadze2020quantum, wang2022supercurrent}, detected with dc transport techniques. Such approaches are typically slow, require macroscopic leads for current injection, and measurements lead to a destruction of the parity state. High-frequency parity and charge readout studies revolved around reflectometry techniques, where the device-under-test was embedded in an impedance-matching tank circuit \cite{schoelkopf1998radio,barthel2010fast} and addressed in the radio frequency \cite{razmadze2019radio} or microwave domain \cite{harabula2017measuring, de2019rapid, de2021rapid}, allowing the detection of parity transitions. Finally, experiments in the high-frequency domain have been, so far, almost exclusively performed in semiconductor nanowires. This is due to the challenging, and sometimes competing, fabrication requirements for integrating microwave hardware on planar hybrid systems~\cite{casparis2018superconducting, PhysRevLett.126.037001, chidambaram2022microwave, harlech2022local, hinderling2023flip}, where fast parity-readout and QP dynamics studies are yet to be demonstrated.

In this Article, we present a novel approach to parity readout in a confined superconducting system. A planar superconducting island is embedded in a superconducting loop, which is inductively coupled to a high $Q$ factor microwave resonator using a flip-chip approach. Parity in the superconducting island directly maps into the direction of the circulating supercurrent in the loop, allowing for fast, absolute and non-destructive parity readout. The parity effect in the island is observed at zero magnetic field concomitantly with the presence of a low-energy bound state in the island. We demonstrate parity detection with signal-to noise ratio $\mathrm{SNR} \approx 3$ for an integration time of $20~\mu$s with detection fidelity for both parity states above 98~$\%$. By continuously monitoring the state of the island, parity switching events are observed, enabling the estimation of parity lifetime approaching millisecond timescales, with no measurable parity transitions within the parity-blockade regime. Combination of inductive readout technology, flip-chip approach and scalable planar heterostructure will pave a path towards coherent addressing and manipulation of large scale hybrid devices.

\begin{figure*}
	\includegraphics[width=\textwidth]{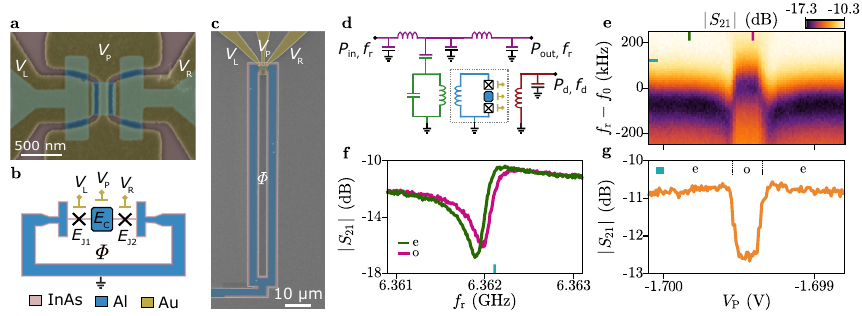}
	\caption{\textbf{Device and parity detection scheme.} \textbf{a}, False-colored scanning electron micrograph of the superconducting island. The Al (blue) was selectively removed to form a 250~nm long and 400~nm wide superconducting island coupled to wide superconducting T-shaped leads. The gate (yellow) with voltage $V_\mathrm{P}$ is used to suppress the parallel transport paths on both sides of the island and for tuning charge occupation. Gates with voltages $V_\mathrm{L}$ and $V_\mathrm{R}$ (yellow) tune the coupling to the leads. \textbf{b}, Schematic representation of the device under study, showing a superconducting island with charging energy $E_\mathrm{C}$ embedded in a superconducting loop tuned by the flux $\it{\Phi}$. Coupling between island and leads is tuned via electrostatically controlled Josephson junctions with Josephson energies $E_\mathrm{J1}$ and $E_\mathrm{J2}$. \textbf{c}, Micrograph of the superconducting loop incorporating the gate-tunable island. \textbf{d}, Schematic of the equivalent parity detection circuit. The readout hang resonator (green) is inductively coupled to the superconducting loop (blue). The readout is performed by probing the complex transmission parameter $S_\mathrm{21}=10\mathrm{log}(P_\mathrm{out}/P_\mathrm{in})$ through a coplanar transmission line (purple) fabricated on the same chip as the resonator and capacitively coupled to the resonator. An additional drive line (red) present on the resonator chip enables two-tone microwave spectroscopy measurements. \textbf{e}, Magnitude of the transmission response $\lvert S_{21} \rvert$ in the Coulomb blockade regime ($V_\mathrm{L}=-1.755$~V and $V_\mathrm{R}=-1.610$~V) as a function of plunger voltage $V_\mathrm{P}$ and offset readout frequency $f_\mathrm{r}-f_\mathrm{0}$, with $f_\mathrm{0}=6.38198$~GHz. The shift of the resonator frequency indicates a transition between the two parity states. \textbf{f},~Linecuts from resonator response in~\textbf{e} at positions indicated by the green and pink markers. An 80~kHz shift of the resonator frequency is present between even (e) and odd (o) parity states. \textbf{g}, A linecut from~\textbf{e} at the turquoise marker position as a function of $V_\mathrm{P}$ emphasizes the change in the resonator magnitude response between even and odd parity state, demonstrating parity detection.}
	\label{fig1}
\end{figure*}

\section*{Device and readout scheme}
Figure~\ref{fig1}a displays a false-colored scanning electron micrograph of the active part of the device. The device consisted of a floating superconducting island (blue) with dimensions of 250~nm $\times$ 400~nm that was coupled to the loop by two semiconducting junctions each of length $\approx 50$~nm. The junctions were controlled with gate voltages $V_\mathrm{L}$ and $V_\mathrm{R}$ (yellow), realizing a tunable coupling to the superconducting leads (blue). The middle gate, energized by voltage $V_\mathrm{P}$, was used both for the depletion of parallel transport paths around the island and for tuning the offset charge on the island. Figures~\ref{fig1}b,c show a schematic and a scanning electron micrograph of the full device, respectively. Relevant device parameters are the charging energy $E_\mathrm{C}$ of the island and the gate-controlled Josephson energies $E_\mathrm{J1}$ and $E_\mathrm{J2}$, describing the energy required to add (remove) a single electron to (from) the island and the coupling between the island and the leads, respectively. To enable flux-biasing, the island was embedded into a grounded superconducting loop (blue). The loop was patterned into the same InAs/Al heterostructure, and was controlled via a magnetic flux $\it{\Phi}$, generated by a home-made superconducting coil mounted on top of the printed circuit board that hosted the device.   

Microwave measurements relied on the flip-chip approach that was recently pioneered in hybrid materials using nanowire-based~\cite{zellekens2022microwave} and planar devices~\cite{hinderling2023flip}. The planar device and microwave components were fabricated on separate chips to achieve high-quality devices and readout resonators ($Q_\mathrm{i}$ = 35000, $f_\mathrm{0}=6.36198$~GHz). The chips were connected with indium bump bonding. While the microwave resonators were inductively coupled to the device via a vacuum gap ($d=5~\mu$m) to sense the supercurrent in the device loop, galvanic connections between both chips were used to tune the electrostatic gates and provide a grounding contact to the loop. 

The simplified microwave measurement scheme is shown in Fig.~\ref{fig1}d. The transmission line (purple) was capacitively coupled to the readout resonator (green), which changed its frequency and loaded quality factor $Q_\mathrm{L}$ due to current circulating in the loop (blue). Parity readout was performed by applying a probe signal $P_\mathrm{in}$ with frequency $f_\mathrm{r}$ into the left port of the transmission line and detecting the response $P_\mathrm{out}$ at the right output port at the same frequency $f_\mathrm{r}$. For two-tone spectroscopy measurements, the additional drive line shown in Fig.~\ref{fig1}d (red) was used to apply a continuous drive tone with power $P_\mathrm{d}$ and frequency $f_\mathrm{d}$ while monitoring the resonator transmission $P_\mathrm{out}$ at frequency $f_\mathrm{r}$. More details about the fabrication, packaging and measurement setup are reported in the Methods section. Data presented here were collected on one device, whereas data from a second device are presented in the Supplementary Section~S1. 

\section*{Parity effect in a superconducting island}

We first discuss the response of the resonator as a function of gate voltage $V_\mathrm{P}$, when $V_\mathrm{L}$ and $V_\mathrm{R}$ were set negative enough to result in a finite charging energy, but still allowed for a supercurrent to flow in the loop. Furthermore, the system was tuned to a regime with strong parity effect. Figure~\ref{fig1}e displays the magnitude of transmission response $\lvert S_{21} \rvert$ as a function of $V_\mathrm{P}$ and offset readout frequency $f_\mathrm{r}-f_\mathrm{0}$, with $f_\mathrm{0}=6.38198$~GHz. As the gate voltage $V_\mathrm{P}$ was increased, a sharp change of the resonator frequency was observed in a narrow gate voltage range around $V_\mathrm{P}=-1.67$~V. In Fig.~\ref{fig1}f, we show linecuts of the resonator magnitude response $\lvert S_{21} \rvert$ as a function of readout frequency $f_\mathrm{r}$ at two voltage $V_\mathrm{P}$ values (dark green and pink markers) indicated in panel e. At these values of $V_\mathrm{P}$ both a shift of the resonator frequency by 80~kHz and an accompanying 12$\%$ change in its linewidth were visible. Remarkably, the frequency shift can be associated to a parity-dependent change in the direction of the circulating current, i.e. in the formation of a $\pi$-junction for an odd-parity ground state. The circulating current results in a sign change of the effective inductance $L= [(2 \pi/\it{\Phi}_\mathrm{0}) \partial I_\mathrm{s}(\varphi)/\partial \varphi]^\mathrm{-1}$, producing a shift in the resonator frequency. Damping is associated to the formation of additional loss channels during the transition through a charge degeneracy point, consistent with the variable load impedance model~\cite{PhysRevResearch.4.013198}. Monitoring the resonator response at fixed $f_\mathrm{r}$ (turquoise marker position in Fig.~\ref{fig1}e) as a function of $V_\mathrm{P}$, as shown in Fig.~\ref{fig1}g, allowed us to track the parity ground state of the superconducting island.

\begin{figure}
	\includegraphics[width=\columnwidth]{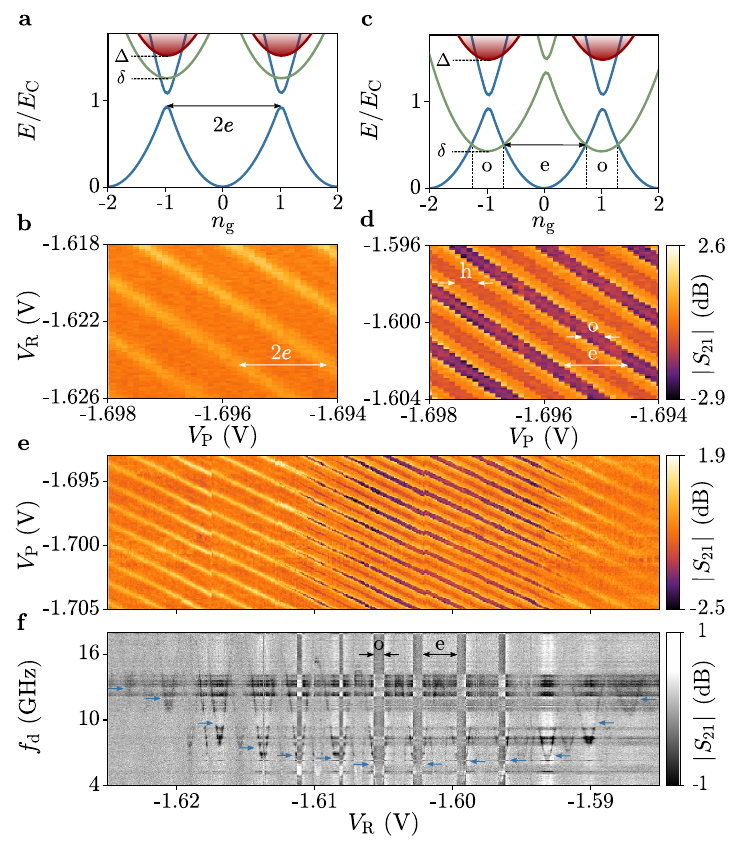}
	\centering
	\caption{\textbf{Charge states of a superconducting island} \textbf{a}, Electrostatic energy diagram of a Coulomb-blockaded superconducting island with charging energy $E_\mathrm{C}$ smaller than a subgap energy state $\delta$ (2$e$ periodic regime). \textbf{b}, Experimentally measured transmission response magnitude $ \lvert S_{21} \rvert$ at $\it{\Phi}=\it{\Phi}_\mathrm{0}/\mathrm{2}$ as a function of plunger gate voltage $V_\mathrm{P}$ and right barrier gate voltage $V_\mathrm{R}$. The 2$e$ periodic Coulomb blockade oscillations observed as a function of both voltages are consistent with the situation depicted in~\textbf{a}. \textbf{c}, Same as \textbf{a} but for $E_\mathrm{C}>\delta$, showing parity effect with alternating even and odd parity valleys. \textbf{d}, Same as \textbf{b} but in a gate regime where the parity effect is present, leading to periodically alternating spacings for even (e) and odd (o) parity sectors. \textbf{e}, Overview of resonator response magnitude $\lvert S_{21} \rvert$ at $\it{\Phi}=\it{\Phi}_\mathrm{0}/\mathrm{2}$ for broad ranges of $V_\mathrm{P}$ and $V_\mathrm{R}$. Tuning of parity effect is observed as a modulation of periodicity between Coulomb blockade peaks as the voltage $V_\mathrm{R}$ is varied. \textbf{f}, Two-tone spectroscopy measurement at $\it{\Phi}=\it{\Phi}_\mathrm{0}/\mathrm{2}$ showing the resonator response magnitude $\lvert S_{21} \rvert$ as a function of drive frequency $f_\mathrm{d}$ and right barrier voltage $V_\mathrm{R}$ in a gate regime where the parity effect is present.}
	\label{fig2}
\end{figure}

The energy diagram of a confined superconducting island in the absence of low energy excitations is shown in Fig.~\ref{fig2}a. The energy of the island is characterized by periodic parabolas describing the even parity states (blue) with charging energy $E_\mathrm{C}$, the induced superconducting gap $\Delta$ (red) and a generic subgap state with energy $\delta$ (green), which is in this case above $E_\mathrm{C}$. In this situation the parity of the island remains even, and sweeping gate voltages changes the charge occupation by two electrons at a time. Experimentally, this was observed in Fig.~\ref{fig2}b as a series of periodic features in transmission response magnitude $\lvert S_{21} \rvert$ as a function of $V_\mathrm{P}$ and $V_\mathrm{R}$, measured at $\it{\Phi}=\it{\Phi}_\mathrm{0}/\mathrm{2}$. The regions where the $\lvert S_{21} \rvert$ signal was enhanced are associated with crossing points of electrostatic energy parabolas in Fig.~\ref{fig2}a. At these points, corresponding to resonant tunnelling of Cooper pairs through the island, a sizeable supercurrent circulated in the loop, whereas between resonances the charge configuration was stable and the supercurrent was blocked. Therefore, $ \lvert S_{21}  \rvert$ monitored at a fixed $f_\mathrm{r}$ allowed to distinguish between the current flowing at charge-state degeneracy and its suppression in the blockade regime.

A significantly different situation is obtained when an excitation with energy $\delta<E_\mathrm{C}$ is present in the spectrum (see Fig.~\ref{fig2}c). In this case, crossing points between the lowest energy parabolas (blue and green) become unevenly spaced, and the superconducting island accepts single electrons. This was measured in Fig.~\ref{fig2}(d). The abrupt and periodic change in color of the $\lvert S_{21} \rvert$ response separates the two parity sectors. The odd parity (o) state is associated to narrow and dark color regions separated by larger bright and wider color regions where parity is even (e). The additional features in the center of the even parity sector (h), running parallel in gate space to the main parity transitions, are associated to transport processes involving excited states of the island. The crossings of the associated higher energy parabolas [higher energy than shown in Fig.~\ref{fig2}c] enable new transport channels and lead to a measurable effect on the current circulating in the loop visible in the even parity valley~\cite{PhysRevLett.118.137701}.

Figure~\ref{fig2}e displays $\lvert S_{21} \rvert$ at $\it{\Phi}=\it{\Phi}_\mathrm{0}/\mathrm{2}$ in a broader parameter space, where transitions between the two regimes discussed above were visible. Results in Fig.~\ref{fig2}e strongly suggest that the charging energy of the superconducting island was generally smaller than the superconducting gap. However, discrete Andreev states could be occasionally tuned to energies $\delta<E_\mathrm{C}$, giving rise to parity transitions. In the case of Fig.~\ref{fig2}(e), such a discrete state was predominantly controlled by $V_{\mathrm{R}}$.

More quantitative insights are gained by performing two-tone spectroscopy measurements, as shown in Fig.~\ref{fig2}f. We observe a driven transition (parabolas indicated by blue arrows), which is continuous and periodic as a function of $V_\mathrm{R}$, with a periodicity consistent with 2$e$ charging. We interpret the driven transition as excitations from the even parity ground state to its excited state (blue curves in Fig.~\ref{fig2}a,c)~\cite{proutski2019broadband}. Here, the minimum transition frequency of each oscillation is consistent with the total Josephson energy of the system $\approx 11$~GHz. As a discrete state is tuned below the charging energy, the ground state parity changes to odd, resulting in sharp color changes over finite regions of $V_\mathrm{R}$. From the transition frequency where ground state transitions first occurred, a charging energy $E_\mathrm{C} \approx 6.7$~GHz was estimated (see Supplementary Section~S2). Horizontal resonances are attributed to unintentional standing waves in the measurement circuit. Additional two-tone spectroscopy measurements are presented in Extended Data Fig.~\ref{s7}.

\begin{figure}
	\includegraphics[width=\columnwidth]{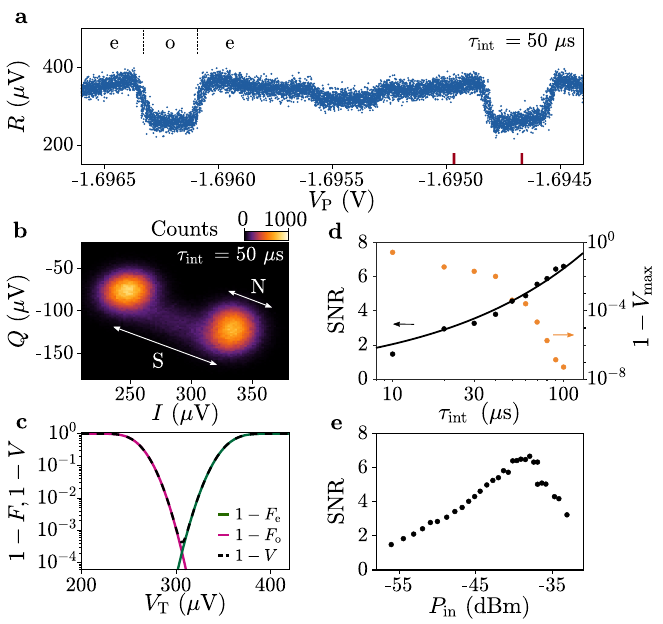}
	\centering
	\caption{\textbf{Signal-to-noise ratio and detection fidelity.} \textbf{a},~Transmission magnitude response $R$ as a function of plunger gate voltage $V_\mathrm{P}$ measured at $V_\mathrm{L}=-1.755$~V and $V_\mathrm{R}=-1.609$~V, using a readout power $P_\mathrm{in}=-44$~dBm (before attenuation) and an integration time $\tau_\mathrm{int}=50~\mu$s. Even~(e) and odd~(o) parity states are identified as different discrete levels in $R$ response. \textbf{b}, Histogram of the resonator magnitude response $R$ decomposed into in-phase ($I$) and quadrature ($Q$) components encoding the two parity states. Signal ($S$) is characterized by the separation between the maxima of two centers while noise ($N$) is responsible for the broadening of the distribution. The data range used for the histogram is between the two red markers in~\textbf{a}. \textbf{c}, Parity readout infidelity for the even (1$-F_\mathrm{e}$, green) and odd (1$-F_\mathrm{o}$, pink) sectors, together with readout invisibility $1-V$ (black dashed) as a function of threshold voltage $V_\mathrm{T}$ separating the two parity sectors for $\tau_\mathrm{int}=50~\mu$s. \textbf{d}, Signal-to-noise ratio (SNR) together with a fit to an uncorrelated noise model (solid black line) and extracted minimum invisibility $1-V_\mathrm{max}$ as a function of integration time $\tau_\mathrm{int}$. \textbf{e},~Dependence of SNR on readout power $P_\mathrm{in}$ (before attenuation) for $\tau_\mathrm{int}=50~\mu$s and $f_{r}=6.36206$~GHz, showing a turnover behavior.} 
	\label{fig3}
\end{figure}

\section*{Estimation of signal-to-noise ratio and readout fidelity}
Fast and continuous monitoring of charge parity provides access to the fast dynamics of the system. The performance of our approach is quantified by the signal-to-noise ratio (SNR) and the measurement fidelity $F$. First, we acquired in-phase ($I$) and quadrature ($Q$) components of the resonator response at fixed readout frequency $f_\mathrm{r}=6.36206$~GHz and calculated the transmission response magnitude ${R=\sqrt{I^2+Q^2}}$ as a function of $V_\mathrm{P}$ for variable data integration times $\tau_\mathrm{int}$. An exemplary $R$ trace measured with $\tau_\mathrm{int}=50~\mu$s is depicted in Fig.~\ref{fig3}a, where two well-distinguishable levels of response are associated with even and odd parity states. Higher order transport processes mentioned above are also visible as a dip in $R$ centered at $V_\mathrm{P}=-1.6955$~V. The $I$ and $Q$ quadrature map of Fig.~\ref{fig3}b provides access to amplitude and phase of the signal, which we next use to estimate the levels of signal ($S$) and noise ($N$). For this, we fit the resulting histogram to a two-dimensional bi-modal Gaussian distribution. The signal level is defined as the distance between centers of the fitted Gaussians, whereas the noise is defined as the standard deviation of the Gaussians responsible for state broadening (see Extended Data Fig.~\ref{s9} for details). 

To calculate the fidelity $F$ of the parity-state assignment, we first project the histogram data along the axis connecting the centers of the two Gaussians to recover a 1D histogram of state distribution. Following this, the fidelities for the even and odd states are defined as ${F_\mathrm{e} = 1 - \int_{V_\mathrm{T}}^{\infty} n_\mathrm{e}(V) dV}$ and ${F_\mathrm{o} = 1 - \int_{-\infty}^{V_\mathrm{T}} n_\mathrm{o}(V) dV}$, respectively \cite{PhysRevLett.103.160503,razmadze2019radio}, where $V_\mathrm{T}$ is the threshold voltage, namely the value of the output voltage that defines the separation between the assignment of data points to even or odd parity sectors, and $n_{e}$ ($n_{o}$) is the probability density of the even (odd) state. The results for the fixed integration time $\tau_\mathrm{int}=50~\mu$s are shown in Fig.~\ref{fig3}c  where both state infidelities 1$-F_\mathrm{e}$ (green) and 1$-F_\mathrm{o}$ (pink) are plotted as a function of $V_\mathrm{T}$. To define the optimum value of $V_\mathrm{T}$ for the two parity states, the visibility is calculated as $V=F_\mathrm{e}+F_\mathrm{o}-1$. Maximizing $V$ allows us to find the optimal regime for $F_\mathrm{e}$ and $F_\mathrm{o}$. In this case the detection is optimized when 1$-V$ reaches a minimum at a threshold voltage $V_\mathrm{T}=306~\mu$V with maximum obtained visibility $V_\mathrm{max}=0.9996$ (three nines). 

The extraction procedure for SNR and $V_\mathrm{max}$ is repeated for multiple $\tau_\mathrm{int}$ values and summarized in Fig.~\ref{fig3}d. Both SNR and $V_\mathrm{max}$ increase with increasing $\tau_\mathrm{int}$. The SNR trace in Fig.~\ref{fig3}d was fitted to an uncorrelated noise model~\cite{barthel2010fast}, ${\mathrm{SNR(\tau_\mathrm{int})}=(S_\mathrm{\tau_\mathrm{int}=1\mu s}/N_\mathrm{\tau_\mathrm{int}=1\mu s})\sqrt{\tau_\mathrm{int} / (1~\mu s)}}$, where $S_\mathrm{\tau_\mathrm{int}=1\mu s}=87~\mu$V is the signal at $\tau_\mathrm{int}=1~\mu$s (used as a fit parameter) and $N_\mathrm{\tau_\mathrm{int}=1\mu s}=135~\mu$V is the noise at $\tau_\mathrm{int}=1~\mu$s, determined by fitting ${N(\tau_\mathrm{int}) \sim 1/\sqrt{\tau_\mathrm{int}}}$. This allowed the estimation of integration time $\tau_\mathrm{min}=2.4~\mu$s for which $\mathrm{SNR}=1$ and parity sensitivity $S=e\sqrt {\tau_\mathrm{min}}=1.5 \times 10^{-3}~e/\sqrt{\mathrm{Hz}}$. The measurement fidelities were found to vary from 85$\%$ at $\tau_\mathrm{int}=10~\mu$s reaching 99.99999$\%$ (seven nines) at $\tau_\mathrm{int}=100~\mu$s. The values obtained here for SNR, fidelity and visibility are comparable to previously reported charge and parity detection studies in hybrid materials \cite{razmadze2019radio, nguyen2022electrostatic}.

Finally, Fig.~\ref{fig3}e displays the SNR dependence on readout power $P_\mathrm{in}$ (before cryogenic attenuation) measured with $\tau_\mathrm{int}=50~\mu$s at a readout frequency ${f_\mathrm{r}=6.36202}$~GHz. The initial increase of $P_\mathrm{in}$ leads to a correlated increase in SNR reaching a maximum of $\mathrm{SNR} \approx 6.7$ at $P_\mathrm{in}=-38$~dBm. Beyond this optimal value, the SNR shows a strong downward trend. This turnover behavior is mainly caused by signal $S$, while noise $N$ is only slightly affected in the $P_\mathrm{in}$ range under study. Here, the signal is set by the frequency shift and damping of the readout resonator, which both decrease for larger $P_\mathrm{in}$ (see Extended Data Fig.~\ref{s6}). For low powers the decreased damping improves the signal, but past the threshold the reduction in frequency shift starts to dominate, which results in smaller signals. We additionally note that the SNR values presented here are likely affected by the $1/f$ noise due to the low sweep rate of the gate voltage $V_\mathrm{P}$, and therefore act as a lower bound for SNR in time-resolved measurements. The SNR dependence on $f_\mathrm{r}$ is shown in Extended Data Fig.~\ref{s8}.

\section*{Time-resolved parity monitoring}

\begin{figure}
	\includegraphics[width=\columnwidth]{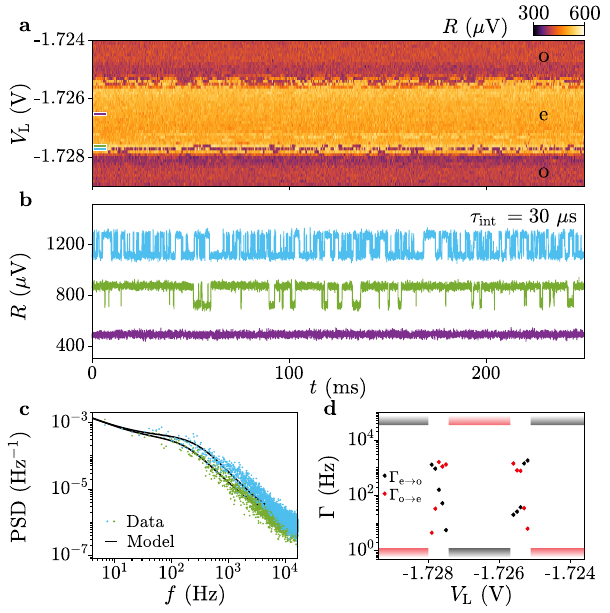}
	\centering
	\caption{\textbf{Continuous parity monitoring.} \textbf{a}, Resonator response magnitude $R$ in the regime where the parity effect is visible as a function of measurement time $t$ and voltage $V_\mathrm{L}$, for $\tau_\mathrm{int}=30~\mu$s. Around the regions where parities are degenerate, the switching rates are enhanced compared to the rates away from degeneracy. \textbf{b}, Blue; time trace of resonator response $R$ close to parity degeneracy, showing parity switching effects detected in real time. Green; similar trace but further away from degeneracy where the switching rate is reduced. Purple; time trace in the blockade regime, where the parity is fixed. \textbf{c}, Welch power spectral density (PSD) as a function of frequency $f$ for the blue and green traces in \textbf{b}, fitted to an asymmetric random telegraph noise model including a $1/f$ noise term. \textbf{d}, Parity switching rates $\Gamma_\mathrm{e \rightarrow o}$ and $\Gamma_\mathrm{o \rightarrow e}$ as a function of voltage $V_\mathrm{L}$ in~\textbf{a}. In the blockade regime the lower and upper bound on $\Gamma_\mathrm{e \rightarrow o}$ and $\Gamma_\mathrm{o \rightarrow e}$, shown as red and black lines, are limited by the length of the total recorded time trace (900~ms) and the integration time ($\tau_\mathrm{int}=30~\mu$s).}
	\label{fig4}
\end{figure}

After assessing good parity sensitivity and detection fidelity, we now study the time-resolved dynamics in the planar island. To detect parity transition events, the resonator response magnitude $R$ was monitored for a time window of 900~ms with $\tau_\mathrm{int}=30~\mu$s for multiple values of $V_\mathrm{L}$. The results are summarized in Fig.~\ref{fig4}a, where we show a subset of that time window spanning 250~ms. Deep in the even parity sector, where the charge configuration is fixed, the resonator response level is stable in time (purple trace in Fig.~\ref{fig4}b). In this case the charging energy of the superconducting island energetically penalizes changes in charge configuration, leading to a stable parity occupation. The step-like transitions observed between the blue and green traces in Fig.~\ref{fig4}b are associated to parity switching events following a random telegraph noise behavior. The frequency of switching events as well as the preferred parity state are tuned by $V_\mathrm{L}$. Additional measurements on gate-dependent parity switching are presented in Extended Data Fig.~\ref{s10}.

The average transition rates in such a two-level system is quantified in the spectral domain using power spectral density (PSD) analysis~\cite{riste2013millisecond, PhysRevLett.120.100502}. For this the time traces were converted to a Welch PSD~\cite{welch1967use}, as shown in Fig.~\ref{fig4}c for the green and blue time traces from Fig.~\ref{fig4}b. The PSD was then fitted with an asymmetric random telegraph noise model including a $1/f$ noise term~\cite{PhysRevB.100.140503}. This allowed us to infer the two parity switching rates $\Gamma_\mathrm{e \rightarrow o}$ and $\Gamma_\mathrm{o \rightarrow e}$ associated to transitions between even-odd and odd-even parity sectors. For the green trace we obtain ${\Gamma_\mathrm{e \rightarrow o}=53 \pm 1}~$Hz and ${\Gamma_\mathrm{o \rightarrow e}=1123 \pm 17}~$Hz. For the blue trace we obtained ${\Gamma_\mathrm{e \rightarrow o}=163 \pm 4}$~Hz and ${\Gamma_\mathrm{o \rightarrow e}=1624 \pm 24}$~Hz, consistent with a reduced asymmetric switching configuration. 

The parity switching rate extraction procedure was repeated at several $V_\mathrm{L}$ values from Fig.~\ref{fig4}a and the results are summarized in Fig.~\ref{fig4}d. In the regions close to parity degeneracy, both rates develop a strong dependence on $V_\mathrm{L}$ and exchange their dominating behavior as a function of detuning on both sides of the degeneracy. In the blockaded regime, the lowest detectable parity switching rate is limited by the length of the total recorded time trace of 900~ms with ${\mathrm{min}(\Gamma_\mathrm{e \rightarrow o})=\mathrm{min}(\Gamma_\mathrm{o \rightarrow e}) =1/(900~\mathrm{ms})}$, whereas the highest detectable rate is limited by the integration time $\tau_\mathrm{int}=30~\mu$s, ${\mathrm{max}(\Gamma_\mathrm{e \rightarrow o})=\mathrm{max}(\Gamma_\mathrm{e \rightarrow o}) = 1/(30~\mu\mathrm{s})}$. The two limits are schematically indicated as red and black regions in Fig.~\ref{fig4}d. The parity switching rates $\Gamma_\mathrm{e \rightarrow o}$, $\Gamma_\mathrm{o \rightarrow e}$ and the corresponding state lifetimes $T_\mathrm{e}$=1/$\Gamma_\mathrm{e \rightarrow o}$, $T_\mathrm{o}$=1/$\Gamma_\mathrm{o \rightarrow e}$ reported here are comparable to values measured for superconducting-semiconducting nanowire qubits and similar hybrid devices \cite{nguyen2022electrostatic, uilhoorn2021quasiparticle, PhysRevB.100.020502}. For the time response shown in blue in Fig.~\ref{fig4}b, we found $T_\mathrm{e} = 6.1 \pm 0.2$~ms and  $T_\mathrm{o}=616 \pm 9~\mu$s, producing a first reported estimate on parity lifetime in a superconducting planar heterostructure. 

\section*{Conclusion}\label{sec13}

We have reported on a flip-chip-based inductive detection scheme of the charge parity in a planar superconducting island. Our method combined the advantages of high-bandwidth readout technology with planar gate-tunable superconductor-semiconductor materials to achieve a high signal-to-noise ratio for non-destructive readout of parity. While the parity effect in our devices was enabled by the low energy excitation in the island, the readout technique is general and does not depend on the presence of such states. The combination of flip-chip approach and inductive readout offers significant improvements whenever the experimental implementation involves circulating currents. Our results open interesting avenues for the realization of theoretical proposals in the field of Andreev spin and topological qubits. Furthermore, it will facilitate the long-range coupling of different qubit subsystems, especially when large-scale devices become available using planar device technology. The long measured parity lifetimes constitute an important advantage in the context of hybrid qubits, where large numbers of gate operations could be performed before the parity information is lost. Additionally, the availability of parity lifetime measurements will allow to constrain theoretical proposals for hybrid and protected qubit designs based on hybrid planar materials.

\section*{Methods}\label{sec11}

\subsection*{Fabrication}
Superconducting islands were fabricated from an InAs/Al heterostructure grown on an InP substrate using electron beam lithography techniques. The semiconductor was etched in a solution of H${}_2$O : C${}_6$H${}_8$O${}_7$ : H${}_3$PO${}_4$ : H${}_2$O${}_2$ with composition 220 : 55 : 3 : 3. The epitaxial Al was etched with a short dip into 50~$^{\circ}$C Transene~D to define the island and the superconducting loop. Under-bump metallization was evaporated on the device contacts \cite{hinderling2023flip}, and the entire chip was covered by a dielectric layer consisting of 3 nm AlO${}_2$ and 15 nm HfO${}_2$, grown by thermal and plasma ALD, respectively. Next, the fine features of the gate electrodes were deposited by electron evaporation of 5~nm Ti and 20~nm Au, followed by the evaporation of thicker gate features (Ti/Al/Ti/Au, 5/250/5/100~nm). For both the inner and outer gates, lift-off was done in DMSO at 120~$^{\circ}$C. A 45~nm layer of Al${}_2$O${}_3$ was then deposited with plasma ALD. With wet etching in BHF 7:1, RIE etching and wet etching in the semiconductor etchant, the semiconductor stack was deep etched in the region between the devices and the bottom edge of the chip, to reduce resonator losses. Finally, the chip was diced into a 3~mm by 3~mm piece.

The readout resonators were fabricated on a high-resistivity Si wafer. First, 200~nm of Nb was sputtered over the entire chip, followed by deposition of a 57~nm AlO${}_2$ hard mask by plasma ALD. Resonator features were defined by RIE in the hard mask. After resist removal in hot DMSO, ICP etching with Cl${}_2$ and Ar was used to pattern the Nb layer. Subsequently, the hard mask was removed with BHF 7:1. Next, the In bumps for flip-chip bonding were patterned by optical lithography and In evaporation. Finally, the wafer was diced into 6~mm by 9~mm chips. The resonator chip is identical to that shown in Fig.~1 of Ref.~\cite{hinderling2023flip}, and includes three $\lambda/4$ CPW resonators capacitively coupled to a common transmission line, drive lines, a ground plane and DC control lines.

\subsection*{Flip-chip technique}
After fabrication of the devices and resonators on separate chips, the two chips were coupled via flip-chip bonding using indium bumps. These bumps provided connections from wire bonding pads on the Si chip to the DC control lines of the gate electrodes and ground of the devices on the InP chip. Furthermore, several more In bumps were included in the final chip to ensure mechanical stability of the structure. The devices were carefully aligned with the shorted ends of the $\lambda/4$ resonators and with the drive lines in order to ensure good coupling. The separation between the two chips was $5~\mathrm{\mu m}$, and was controlled by the size of the In bumps as well as the temperature and force used during the flip-chip bonding. To minimize dielectric losses, the device chip was considerably smaller than the resonator chip and was positioned to minimize its overlap with the resonators.

\subsection*{Measurement setup}
Measurements were performed in a BlueFors cryogen-free dilution refrigerator with a mixing chamber base temperature of 9~mK. The wiring of the dilution refrigerator is illustrated in Extended Data Fig.~\ref{s1}. The measurements depicted in Figs.~\ref{fig1} and \ref{fig2} of the Main Text were taken with a Keysight VNA B2911A vector network analyser, while those in Figs.~\ref{fig3} and \ref{fig4} were acquired with a Zurich Instruments SHFQA 8.5 GHZ Quantum Analyzer. The readout tone with frequency $f_\mathrm{r}$ was applied at port $P_\mathrm{in}$. For two-tone spectroscopy measurements (see Fig.~\ref{fig2}f), a continuous drive tone with frequency $f_\mathrm{d}$ was applied at port $P_\mathrm{out}$ in Extended Data Fig.~\ref{s1}. Drive signals were attenuated by 66~dB and 39~dB, respectively, at different temperature stages of the dilution refrigerator. The signal transmitted through the readout transmission line passed through a DC block followed by a circulator, a dual isolator and an rf low-pass filter. The signal was then amplified with a cryogenic high electron mobility transistor amplifier at 4~K and a room-temperature amplifier before being detected at port labelled as $P_\mathrm{out}$. Gates were controlled with QDevil digital-to-analog converter DC voltage sources with 19~$\mu$V resolution. The DC lines were filtered using a home-made low-pass filter at room temperature, QDevil rf and RC filters at base temperature, and an RC filter on the printed circuit board onto which the chip was mounted. The magnetic flux in the loop was applied via a home-made superconducting coil mounted on top of the printed circuit board. 

\subsection*{Device tuneup}
For each device, the superconducting island was electrically defined by tuning the left and right island barrier gates with voltages $V_\mathrm{L}$ and $V_\mathrm{R}$, respectively as well as the plunger gate with voltage $V_\mathrm{P}$. Extended Data Fig.~\ref{s2}a shows flux dependence measured with all gate voltages grounded, resulting in a large circulating current. Extended Data Fig.~\ref{s2}b shows flux dependence measured with all gates set to large negative voltages, resulting in suppression of the circulating current and absence of a resonator response. Dependences of $\lvert S_{21} \rvert$ on various gate voltages and in different gate configurations are shown in Extended Data Figs.~\ref{s3}a-e. The operating regime was found first by setting $V_\mathrm{P}=-1.5$~V and $\it{\Phi}=\it{\Phi}_\mathrm{0}/\mathrm{2}$ and by mapping $\lvert S_{21} \rvert$ as a function of $V_\mathrm{L}$ and $V_\mathrm{R}$. A corner diagram as shown in Extended Data Fig.~\ref{s3}f was recorded, indicative of quantum dot formation. Here, fine tuning of each gate led to the electrostatic regime presented in the Main Text. Additional electrostatic configurations, characterized by different coupling to the leads, are presented in Extended Data Figs.~\ref{s4} and~\ref{s5}. Power dependent measurements of Coulomb blockade transitions and resonator response are shown in Extended Data Fig.~\ref{s6}.

\subsection*{Compensated resonator measurements}
Measurements in Figs.~\ref{fig2}b,d,e were taken with resonator compensation. Each time the slow axis variable was swept to the next value, $ \lvert S_{21}  \rvert$ as a function of readout frequency $f_\mathrm{r}$ was recorded, similar to Fig.~\ref{fig1}f, and the operating frequency was offset by 90~kHz with respect to the frequency of the minimum. During post-processing, the median of the resonator response along the fast axis was subtracted from all data points, resulting in maps of compensated resonator response as a function of two gate voltages. The maps were taken with $P_\mathrm{in} = -44$~dBm, integration time of $420~\mu$s and 15~averages. 

\subsection*{Signal-to-noise ratio, Fidelity and Visibility calculation}
For the signal-to-noise (SNR) calculation, the in-phase ($I$) and quadrature ($Q$) components of the readout resonator was recorded as a function of the plunger gate voltage $V_\mathrm{P}$ with different integration times $\tau_\mathrm{int}$ per data point, ranging from 10 to 100~$\mu$s. The readout frequency was fixed at $f_\mathrm{r}=6.36206$~GHz and the power at $P_\mathrm{in}=-44$~dBm. At each value of the plunger gate voltage, 5000 measurements of $I$ and $Q$ were acquired (10,000 for $\tau_\mathrm{int}=10~\mu$s). Plotting the response magnitude $R=\sqrt{I^2+Q^2}$ (see Fig.~\ref{fig3}a) as a function of $V_\mathrm{P}$ reveals the two parity states with the even state influenced by higher order transport processes. In addition, the amount of recorded data as a function of $V_\mathrm{P}$ is much larger in the even parity state than in the odd state. Therefore, we selected a data set with 120 $V_\mathrm{P}$ values around a parity transition to have similar amount of data in the even and odd parity state. The $I$ and $Q$ values were then plotted as a histogram with bin size (Fig.~\ref{fig3}b). The histogram was fitted with a sum of two 2D Gaussians, Eq.~\ref{eq:s1}

\begin{multline}
	A_1 \cdot e^{-(((b_{I1}-x)/c_{I1})^2+((b_{Q1}-y)/c_{Q1})^2)/2} \\
      + A_2 \cdot e^{-(((b_{I2}-x)/c_{I2})^2+((b_{Q2}-y)/c_{Q2})^2)/2}
	\label{eq:s1}
\end{multline}

where $A_i$ is the height of the Gaussian, $b_{Ii}$ ($b_{Qi}$) the center point and $c_{Ii}$ ($c_{Qi}$) the width of the Gaussian on the $I$-axis ($Q$-axis), for $i=1,2$. The distance between the center points of the two Gaussians defined the strength of the signal $S$, whereas the quadratic sum of the average widths of the Gaussians, projected along $S$ and perpendicular to $S$ using the polar form of ellipse equation, defined the noise $N$. SNR was then defined as $S/N$ and is given by~Eq.~\ref{eq:s2}

\begin{equation}
	\begin{split}
	SNR & = S/N \\
	S & = \sqrt{(b_{I1}-b_{I2})^2 + (b_{Q1}-b_{Q2})^2} \\
	N & = \sqrt{((|c_{\parallel S1}| + |c_{\perp S1}|)/2)^2 + ((|c_{\parallel S2}| + |c_{\perp S2}|)/2)^2}
	\end{split}
	\label{eq:s2}
\end{equation}

where $c_{\parallel Si}$ and $c_{\perp Si}$ are the widths of the Gaussian along and perpendicular to $S$ for $i=1,2$. The SNR for all integration times is plotted in Fig.~\ref{fig3}d and it was fitted to $\mathrm{SNR(\tau_\mathrm{int})}=(S_\mathrm{\tau_\mathrm{int}=1\mu s}/N_\mathrm{\tau_\mathrm{int}=1\mu s})\sqrt{\tau_\mathrm{int}/ (1~\mu s)}$, where $S_\mathrm{\tau_\mathrm{int}=1\mu s}$ represents the signal at $\tau_\mathrm{int}=1~\mu$s and was used as a fit parameter and $N_\mathrm{\tau_\mathrm{int}=1\mu s}=135~\mu$V is the noise at $\tau_\mathrm{int}=1~\mu$s and determined by fitting $N(\tau_\mathrm{int}) \sim 1/\sqrt{\tau_\mathrm{int}}$ assuming $S$ to be independent of $\tau_\mathrm{int}$~\cite{barthel2010fast}.

Readout visibility is defined as $V = F_\mathrm{e} + F_\mathrm{o} - 1$, where $F_\mathrm{e}$ ($F_\mathrm{o}$) is the fidelity of the even (odd) state, respectively. These fidelities are defined as $F_\mathrm{e} = 1 - \int_{V_\mathrm{T}}^{\infty} n_\mathrm{e}(V) dV$ and $F_\mathrm{o} = 1 - \int_{-\infty}^{V_\mathrm{T}} n_\mathrm{o}(V) dV$, where $n_\mathrm{e}$ ($n_\mathrm{o}$) is the probability density of the even (odd) state~\cite{PhysRevLett.103.160503}. The fidelity of a state describes the probability of an assignment of resonator response to the correct parity sector, where $V_\mathrm{T}$ is the voltage threshold, describing the value of transmission output voltage dividing the assignment to two parity sectors. Optimal assignment of $V_\mathrm{T}$ for detection is the value for which the Visibility $V$ is maximized. Maximising $V$ simultaneously also maximizes both detection fidelities $F_\mathrm{e}$ and $F_\mathrm{o}$. To calculate the $F_\mathrm{e}$, $F_\mathrm{o}$ and $V$ a linecut of the data in the IQ plane (see Extended Data Fig.~\ref{s9}a,d) was taken along signal $S$ axis (100 points). The resulting 1D histogram as a function of the voltage $V_\parallel$ was normalized by dividing the data by the sum of all counts (see Extended Data Fig.~\ref{s9}b,e). The normalized histogram was fitted with two 1D Gaussians given by~Eq.~\ref{eq:s3}

\begin{equation}
	A_1 \cdot e^{-((b_1-x)/c_1)^2/2} + A_2 \cdot e^{-((b_2-x)/c_2)^2/2}
	\label{eq:s3}
\end{equation}

where $A_i$ is the height of the Gaussian, $b_{i}$ the center point and $c_{i}$ the width of the Gaussian for $i=1,2$. The integral in $F_\mathrm{e}$ ($F_\mathrm{o}$) was found through numerical integration of the corresponding 1D Gaussian with the composite trapezoidal rule using the python function numpy.trapz. We started (stopped) the integral by the threshold voltage $V_\mathrm{T}$. Finally, the integral was normalized because the histogram data was normalized to the sum of both 1D Gaussians, resulting in~Eq.~\ref{eq:s4}

\begin{equation}
	\begin{split}
	F_\mathrm{e} = 1 - \int_{V_\mathrm{T}}^{\infty} A_2 \cdot e^{-((b_2-V_\parallel)/c_2)^2/2} dV_\parallel \\
	/ \int_{-\infty}^{\infty} A_2 \cdot e^{-((b_2-V_\parallel)/c_2)^2/2} dV_\parallel \\
	 F_\mathrm{o} = 1 - \int_{-\infty}^{V_\mathrm{T}} A_1 \cdot e^{-((b_1-V_\parallel)/c_1)^2/2} dV_\parallel \\
	 / \int_{-\infty}^{\infty} A_1 \cdot e^{-((b_1-V_\parallel)/c_1)^2/2} dV_\parallel.
	\end{split}
	\label{eq:s4}
\end{equation}

Fidelities and visibility as a function of $V_\mathrm{T}$ are plotted in Fig.~\ref{fig3}c and Extended Data Fig. ~\ref{s9}c,f. 

\subsection*{Parity switching rate estimation}

To estimate the parity switching rate we acquired the time trace of the transmission response magnitude $R$ for different left barrier gate voltages $V_\mathrm{L}$, while the other gate voltages were fixed at $V_\mathrm{R} =-1.59800$~V and $V_\mathrm{P} =-1.69219$~V. The readout frequency was $f_\mathrm{r}=6.36206$~GHz and the power was $P_\mathrm{in}=-40.5$~dBm. At each value of $V_\mathrm{L}$, 30001 data points were recorded. The integration time was set to 30~$\mu$s, resulting in a 900~ms long time trace (neglecting the sampling rate of 0.5~ns). The data of the time trace at $V_\mathrm{L}=-1.7277$~V (blue time trace in Fig.~\ref{fig4}b), which showed parity switching between even and odd parity state due to random telegraph noise, was converted into a histogram distributed into 20 bins along $R$. That histogram was fitted with two 1D Gaussians as described in Eq.~\ref{eq:s3}. All time traces were normalized based on this fit parameters so that the transmission response magnitude $R$ became 1 and 0 for the even and odd parity state, respectively. 

Subsequently, the time traces were used to calculate the Welch power spectral density (PSD). The calculation was done using the python function scipy.signal.welch with sampling frequency $f_s = 1/(30~\mu\mathrm{s})$ and with segment length 5001. The resulting PSD was fitted to an asymmetric random telegraph noise model including $1/f$ noise in the system using Eq.~\ref{eq:s5} to estimate the transition rates from the even to odd $\Gamma_{e \rightarrow o}$ and from the odd to even $\Gamma_{o \rightarrow e}$ parity state~\cite{riste2013millisecond, PhysRevLett.120.100502}.

\begin{multline}
	\mathrm{PSD}(f) = \frac{8 \Gamma_{e \rightarrow o} \Gamma_{o \rightarrow e}}{(\Gamma_{e \rightarrow o} + \Gamma_{o \rightarrow e})((\Gamma_{e \rightarrow o} + \Gamma_{o \rightarrow e})^2 + (2\pi f)^2)} \\
	+ B/f+C 
	\label{eq:s5}
\end{multline}

where $B$ represents the influence of $1/f$ noise in the system and $C$ is a constant describing the detection limit. The detection limit is given by the SNR, and we start to approach it using 30~$\mu$s integration time, which can be seen by the flattening PSD curve at high frequencies. Therefore, $C$ was determined as the average of the PSD for the highest 100 frequency points. Since $1/f$ noise influences the slope of the PSD curve, the parameter $B$ was estimated empirically for every time trace with  $2 \cdot 10^{-6}$ times its PSD range. The transition rates  $\Gamma_{e \rightarrow o}$ and $\Gamma_{o \rightarrow e}$ were used as free fit parameters. Since~Eq.~\ref{eq:s5} is symmetric with respect to $\Gamma_{e \rightarrow o}$ and $\Gamma_{o \rightarrow e}$, the initial guess determined which of the two quantities was larger. 

The maximum and minimum detectable transition rates were set by the total time of the time trace and the integration time, respectively. Therefore, the lower bound of the rates is $1/(900~\mathrm{ms})$ (total measurement time) and the upper bound is $1/(30~\mu\mathrm{s})$ (integration time). For resonator response levels that were stable within the measured time, as the purple time trace in Fig.~\ref{fig4}b, its  $\Gamma_{e \rightarrow o}$ and $\Gamma_{o \rightarrow e}$ were assigned to the boundary rates.

\section*{Data availability}
Data presented in this work will be available on Zenodo. The data that support the findings of this study are available upon reasonable request from the corresponding author.

\bibliography{bibliography.bib}

\section*{Acknowledgments}
We thank the Cleanroom Operations Team of the Binnig and Rohrer Nanotechnology Center (BRNC) for their help and support. We are greatful to R.~\v{Z}itko and L.~Pave\v{s}i\'c for helpful discussions. W.W. acknowledges support from the Swiss National Science Foundation (Grant No. 200020 207538). F. N. acknowledges support from the European Research Council (Grant No. 804273) and the Swiss National Science Foundation (Grant No. 200021 201082). 

\section*{Author contributions}
F.N. and D.S. conceived the experiments. E.C., F.K., R.S. and W.W. developed and provided the heterostructure material. M.H. and S.C.t.K. designed the devices. M.H. fabricated the sample and S.P. performed the indium evaporation. M.H. performed the electrical measurements with contributions from D.S. and S.C.t.K.. M.H., S.C.t.K. and D.S. analyzed and interpreted the data with inputs from F.N., D.Z.H. and M.C. M.H., D.S. and S.C.t.K. prepared the manuscript with feedback from all the authors.

\newpage
\setcounter{section}{0}
\onecolumngrid

\renewcommand{\figurename}{\textbf{Extended Data Fig.}}
\newcounter{myc}
\renewcommand{\thefigure}{\arabic{myc}}

\newcounter{mye}
\renewcommand{\theequation}{S.\arabic{mye}}

\newpage	
\setlength{\parskip}{0pt}

\section*{Supplementary Information}

\subsection{Extended Data Figures}

\setcounter{myc}{1}
\begin{figure*}[h]
	\includegraphics[width=\textwidth]{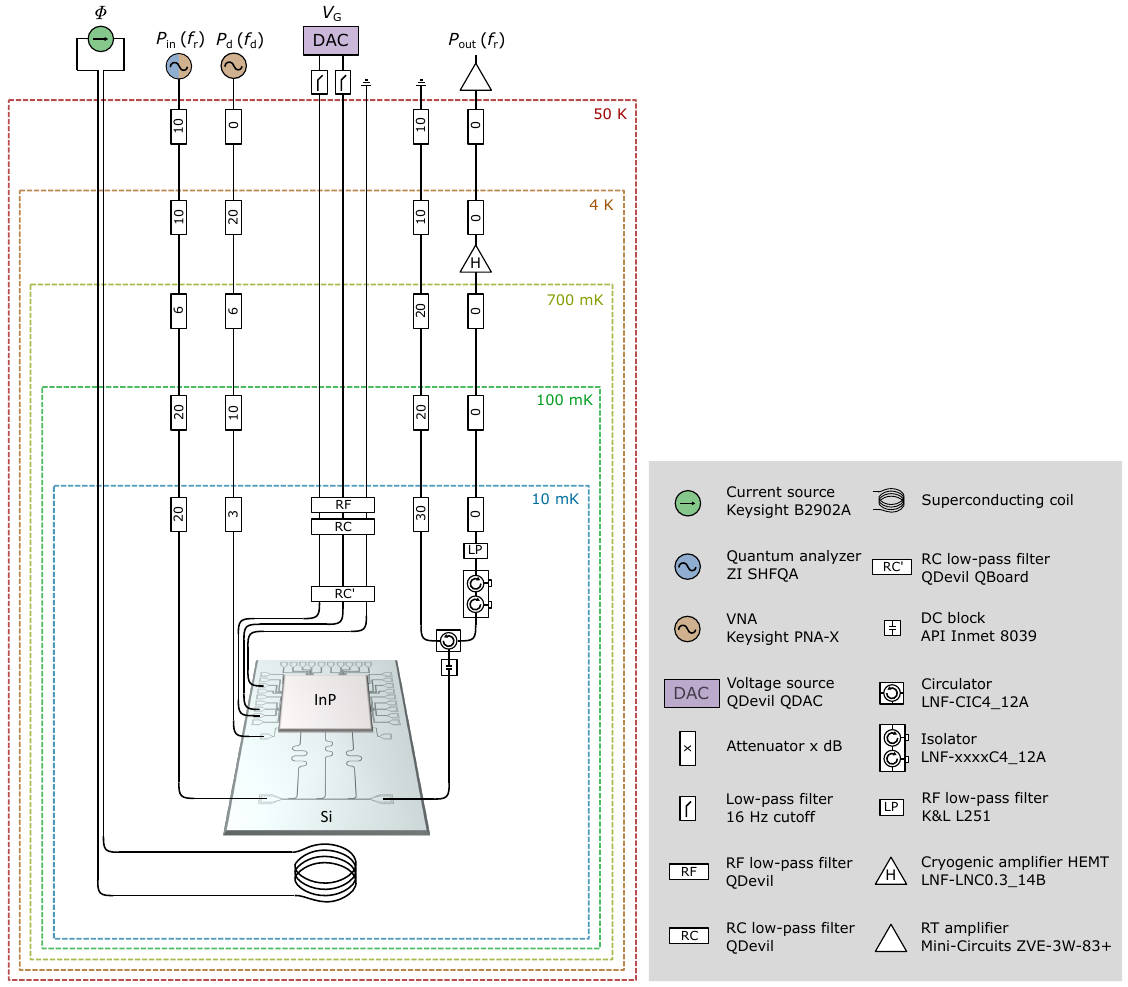}
	\centering
	\caption{\textbf{Schematics of the measurement setup.} The resonator readout and device drive tones with frequencies $f_\mathrm{r}$ and $f_\mathrm{d}$ were applied using a vector network analyzer (VNA) or Quantum Analyser, corresponding to $P_\mathrm{in}$ and $P_\mathrm{d}$, respectively. After 66~dB and 39~dB attenuation of $P_\mathrm{in}$ and $P_\mathrm{d}$ at different temperature stages of the dilution refrigerator, the signals reached the resonator chip. After the transmission, the readout signal passed through a DC block followed by a circulator, a dual isolator and an rf low-pass filter. The signal was amplified with a cryogenic high electron mobility transistor amplifier at 4 K and a room-temperature amplifier, before being detected ($P_\mathrm{out}$) by the VNA or Quantum Analyser. An extra rf line, which could be used for reflectometry measurements, stayed grounded during the experiments. The gates were controlled with digital-to-analog converter DC voltage sources. The DC lines were filtered using a home-made low-pass filter at room temperature, QDevil rf and RC filters at base temperature and an RC filter on the printed circuit board. The applied DC signal reached the device via the resonator chip through the indium bumps. The external flux was applied by sourcing a DC current through a coil, mounted on top of the sample space. There was no additional shielding of the sample space because the refrigerator was equipped with a vector magnet which, despite not being utilised for the experiments.}
	\label{s1}
\end{figure*}

\setcounter{myc}{2}
\begin{figure}[H]
	\includegraphics[width=0.5\columnwidth]{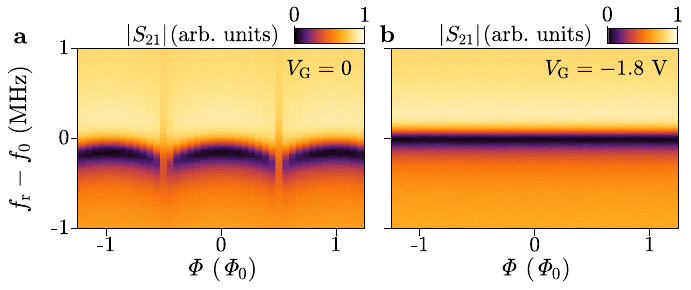}
	\centering
	\caption{\textbf{Flux dependence in Device~1.} \textbf{a}, Normalized resonator magnitude response $\lvert S_{21}  \rvert$ as a function of flux $\it{\Phi}$ and offset readout frequency $f_\mathrm{r}-f_\mathrm{0}$ at $V_\mathrm{G} \equiv V_\mathrm{P}=V_\mathrm{L}=V_\mathrm{R} = 0$. The distinct modulation of $\lvert S_{21} \rvert$ with respect to $\it{\Phi}$ reveals the coupling between resonator and device and indicates a large supercurrent in the device loop carried by many Andreev bound states located between the two superconducting leads~\cite{hinderling2023flip}. \textbf{b}, Same as~\textbf{a} but at $V_\mathrm{G} \equiv V_\mathrm{P}=V_\mathrm{L}=V_\mathrm{R} = -1.8$ V. Depletion of the InAs quantum well between the leads suppresses the supercurrent.}
	\label{s2}
\end{figure}

\setcounter{myc}{3}
\begin{figure}[H]
	\includegraphics[width=0.5\columnwidth]{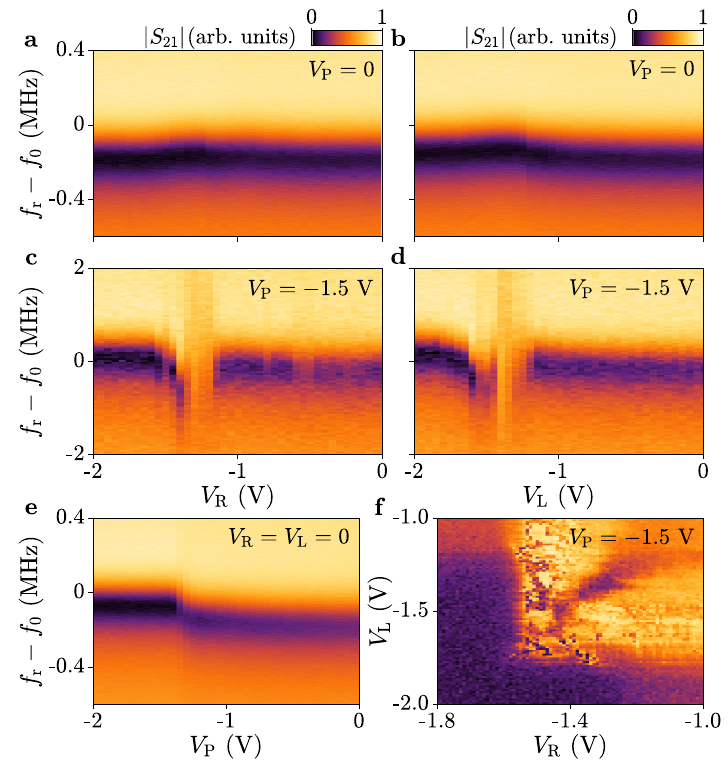}
	\centering
	\caption{\textbf{Tuneup of Device~1.} \textbf{a}, Normalized resonator magnitude response $\lvert S_{21} \rvert$ as a function of offset readout frequency $f_\mathrm{r}-f_\mathrm{0}$ and barrier voltage $V_\mathrm{R}$ at $\it{\Phi}=\it{\Phi}_\mathrm{0}/\mathrm{2}$, $V_\mathrm{P}=V_\mathrm{L}=0$. \textbf{b}, Same as~\textbf{a} but sweeping $V_\mathrm{L}$ and keeping $V_\mathrm{P}=V_\mathrm{R}=0$. Transport channels parallel to the island limited the effect of the barrier gates $V_\mathrm{R}$ and $V_\mathrm{L}$, respectively. \textbf{c}, Same as \textbf{a} but at $V_\mathrm{P}=-1.5$~V. \textbf{d}, Same as \textbf{b} but at $V_\mathrm{P}=-1.5$~V. Suppression of parallel conductance channels increased the influence of the barrier gates on $\lvert S_{21} \rvert$. The coupling between island and lead decreases by reducing the density of Andreev bound states between them. This is observed for $V_\mathrm{R}, V_\mathrm{L} < -1$~V. Appearing resonant transport damped the resonator and caused a shift in its resonance frequency which both decrease for more negative barrier gate voltages. \textbf{e}, Resonator magnitude response $\lvert S_{21} \rvert$ as a function of offset readout frequency $f_\mathrm{r}-f_\mathrm{0}$ and plunger voltage $V_\mathrm{P}$ at $\it{\Phi}_\mathrm{0}/\mathrm{2}$, $V_\mathrm{L}=V_\mathrm{R}=0$. \textbf{f}, Resonator magnitude response $ \lvert S_{21} \rvert$ as a function of barrier gate voltages $V_\mathrm{L}$ and $V_\mathrm{R}$ at $V_\mathrm{P}=-1.5$~V and $f_\mathrm{r}=6.362$~GHz. The corner at $V_\mathrm{L} \approx -1.7$~V and $V_\mathrm{R} \approx -1.5$~V defines the gate range where Coulomb blockad was observed.}
	\label{s3}
\end{figure}

\setcounter{myc}{4}
\begin{figure}[H]
	\includegraphics[width=0.5\columnwidth]{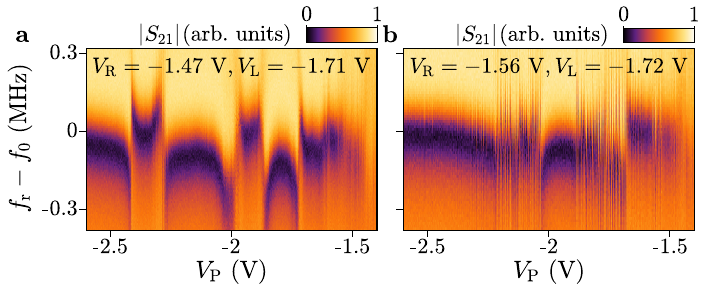}
	\centering
	\caption{\textbf{Flux dependence in Device~1 in different gate settings.} \textbf{a}, Normalized resonator magnitude response $\lvert S_{21} \rvert$ as a function of offset readout frequency $f_\mathrm{r}-f_\mathrm{0}$ and plunger gate voltage $V_\mathrm{P}$ at $V_\mathrm{R}=-1.47$~V, $V_\mathrm{L}=-1.71$~V and $\it{\Phi}=\it{\Phi}_\mathrm{0}/\mathrm{2}$. The barrier gates put the system in a state where there are isolated Andreev bound states (ABSs), but the island is still strongly coupled to the leads. The energy of these bound states is tuned by $V_\mathrm{P}$, leading to anti-crossings when the energy of the ABSs decreases below the resonator frequency consistent with a transmission close to one. \textbf{b}, Same as \textbf{a} but at $V_\mathrm{L}=-1.56$~V and $V_\mathrm{L}=-1.72$~V. The weaker coupling of the island to the leads gives sharp transitions in the regime where ABS energies are below the resonator frequency. We assign these sharp lines to Coulomb transitions between even and odd parity states.}
	\label{s4}
\end{figure}

\setcounter{myc}{5}
\begin{figure}[H]
	\includegraphics[width=0.5\columnwidth]{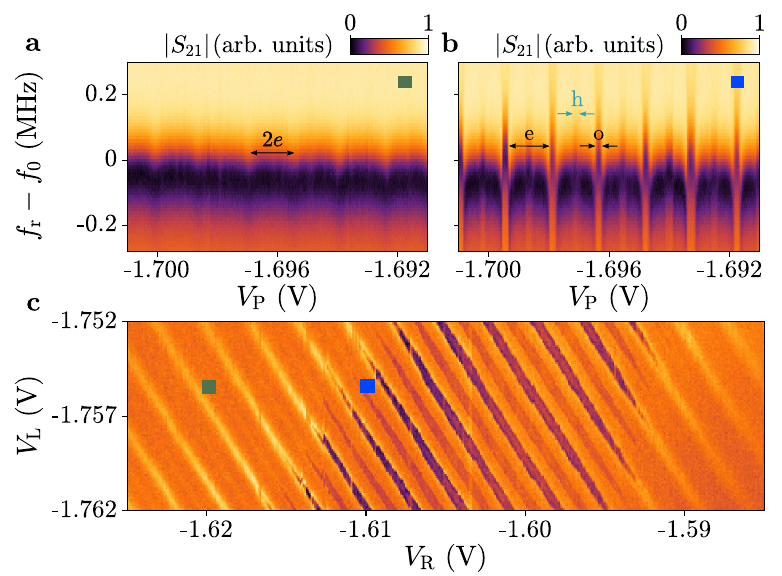}
	\centering
	\caption{\textbf{Additional measurement data on Device~1.} \textbf{a}, Normalized resonator magnitude response $\lvert S_{21} \rvert$ as a function of offset readout frequency $f_\mathrm{r}-f_\mathrm{0}$ and plunger gate voltage $V_\mathrm{P}$ at $V_\mathrm{R}=-1.610$~V, $V_\mathrm{L}=-1.755$~V and $\it{\Phi}=\it{\Phi}_\mathrm{0}/\mathrm{2}$. The 2$e$ periodic Coulomb blockade oscillations are consistent with the charging energy of the island being lower than the bound state energy ($E_\mathrm{C}<\delta$). At Coulomb resonances, a shift in the resonator frequency is observed due to enhanced circulating current in the device loop, which changes the inductance of the device. \textbf{b}, Same as \textbf{a}, but at  $V_\mathrm{R}=-1.620$~V. Here, $\delta < E_\mathrm{C}$ that allows the number of quasiparticles on the superconducting island to be even or odd. The inductance of the even (e) and odd (o) parity states are different resulting in distinct resonator responses. A slight modification of the even state is associated to higher order transport processes~(h)~\cite{PhysRevLett.118.137701}. In the future, such quasiparticle poisoning from strong coupling to the leads could be suppressed with larger $E_\mathrm{C}$. When the charging energy of the superconducting island is larger, a larger energy of the bound state is sufficient to enter the even-odd parity regime, resulting in less coupling between the island and the leads. In that case, stronger coupling between resonator and device, resonators with higher $Q$-factors, or an improvement in the readout setup is required to achieve the same readout performance. \textbf{c}, Normalized resonator magnitude response $\lvert S_{21} \rvert$ as a function of barrier gate voltages $V_\mathrm{L}$ and $V_\mathrm{R}$ at $V_\mathrm{P}=-1.7$~V. The gate-voltage map shows Coulomb blockade oscillations because both barrier gates are able to tune the chemical potential of the superconducting island, but with less efficiency compared to $V_\mathrm{P}$. The bound state energy $\delta$ is mainly modulated by $V_\mathrm{R}$. If $\delta$ becomes lower than $E_\mathrm{C}$, a transition from a 2$e$ periodic to an even-odd regime happens. The odd parity ground state emerge via splitting of the Coulomb resonance peak.}
	\label{s5}
\end{figure}

\setcounter{myc}{6}
\begin{figure}[H]
	\includegraphics[width=0.7\columnwidth]{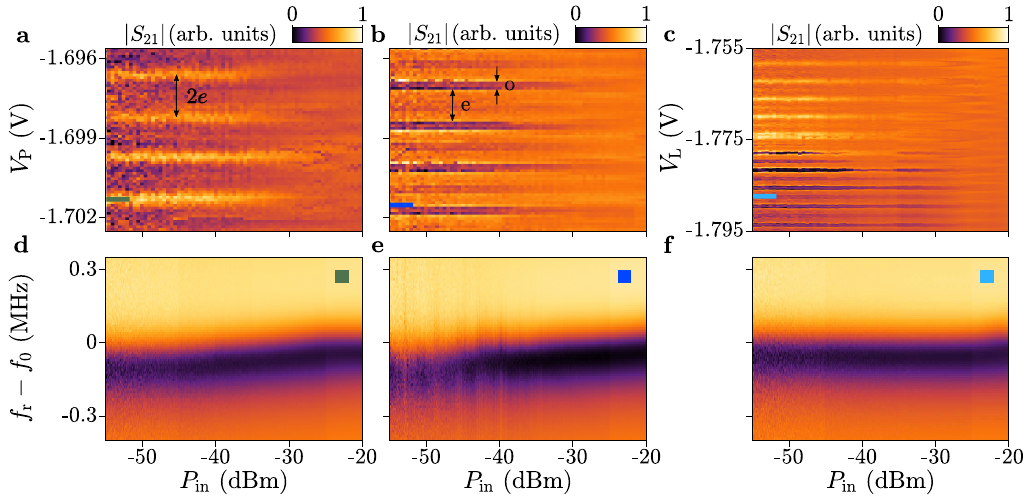}
	\centering
	\caption{\textbf{Power dependence measurements of Device~1.} \textbf{a}, Normalized resonator magnitude response $\lvert S_{21}\rvert$ as a function of input readout power $P_\mathrm{in}$ and plunger gate voltage $V_\mathrm{P}$ in the regime with 2$e$ periodic Coulomb blockade oscillations ($V_\mathrm{L}=-1.760$~V, $V_\mathrm{R}=-1.623$~V) at $\it{\Phi}=\it{\Phi}_\mathrm{0}/\mathrm{2}$. \textbf{b}, Same as \textbf{a}, but in the even-odd regime ($V_\mathrm{L}=-1.795$~V, $V_\mathrm{R}=-1.623$~V). \textbf{c}, Normalized resonator magnitude response $\lvert S_{21}\rvert$ as a function of input readout power $P_\mathrm{in}$ and barrier gate voltage $V_\mathrm{L}$ at $\it{\Phi}=\it{\Phi}_\mathrm{0}/\mathrm{2}$. \textbf{d}, Normalized resonator magnitude response  $\lvert S_{21}\rvert$ as a function of offset readout frequency $f_\mathrm{r}-f_\mathrm{0}$ and input readout power $P_\mathrm{in}$ on a Coulomb resonance peak indicated by the green marker in \textbf{a}. At low input powers the resonator frequency is independent of $P_\mathrm{in}$, but the noise increases for decreasing $P_\mathrm{in}$. Above $P_\mathrm{in}=-44$~dBm the resonator frequency shift decreases for increasing $P_\mathrm{in}$. \textbf{e}, Same as \textbf{d}, but on a transition between even and odd parity state indicated by the blue marker in \textbf{b}. We observe a frequency shift and damping of the readout resonator as a function of $P_\mathrm{in}$. The damping increases significantly for $P_\mathrm{in} < -38$~dBm. \textbf{f}, Same as \textbf{e}, but in a Coulomb blockade regime indicated by the turquoise marker in \textbf{c}. Except for large input powers the resonator frequency stays constant while the noise increases for decreasing $P_\mathrm{in}$.}
	\label{s6}
\end{figure}

\setcounter{myc}{7}
\begin{figure}[H]
	\includegraphics[width=0.5\columnwidth]{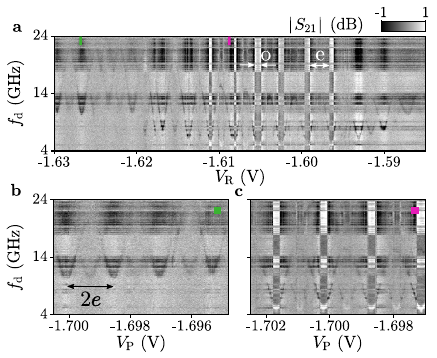}
	\centering
	\caption{\textbf{Additional two-tone spectroscopy measurements in Device~1.} \textbf{a}, Resonator response magnitude $\lvert S_{21} \rvert$ as a function of drive frequency $f_\mathrm{d}$ and right barrier gate voltage $V_\mathrm{R}$ at $V_\mathrm{L}=-1.755$~V, $V_\mathrm{P}=-1.6955$~V and $\it{\Phi}=\it{\Phi}_\mathrm{0}/\mathrm{2}$. While the horizontal lines are associated to unintentional resonances in the measurement circuit, the vertical stripes with opposite magnitude shift (black and white inverted) indicate the odd ground state. The continuous parabola-curved lines, which are periodic with $V_\mathrm{R}$, are associated to driven transitions from the even parity ground state to its excited state. Deep in the 2$e$ periodic regime ($V_\mathrm{R}<-1.62$~V where $E_\mathrm{C} \ll \delta$) the transition parabolas have constant minimum transition frequency $f_\mathrm{d} \approx 11$~GHz corresponding to the total Josephson energy of the device. The transition parabolas get modulated in the even-odd regime and close to it ($V_\mathrm{R}>-1.62$~V) because the barrier gate voltage tunes the gate induced charge of the island, the bound state energy $\delta$ and the coupling to the right lead. The explanation of the modulated transitions goes beyond the simple picture in Fig.~\ref{fig2} of the Main Text. \textbf{b}, Resonator response magnitude $\lvert S_{21} \rvert$ as a function of drive frequency $f_\mathrm{d}$ and plunger gate voltage $V_\mathrm{P}$ at $V_\mathrm{R}=-1.627$~V (green marker position in~\textbf{a}). The spectroscopy shows 2$e$ periodic transition parabolas because $V_\mathrm{P}$ tunes only the gate induced charge of the island. Weaker transitions shifted by 1$e$ are assigned to higher order transport processes~\cite{PhysRevLett.118.137701}. \textbf{c}, Same as \textbf{b} but in the even-odd regime at the pink marker position in~\textbf{a}.}
	\label{s7}
\end{figure}

\setcounter{myc}{8}
\begin{figure}[H]
	\includegraphics[width=0.35\columnwidth]{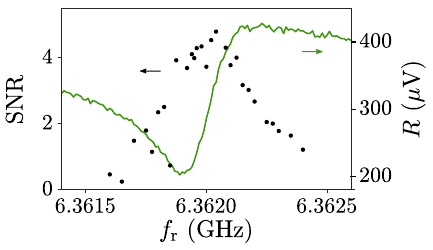}
	\centering
	\caption{\textbf{Frequency-dependent signal-to-noise ratio.} Signal-to-noise ratio (SNR) as a function of readout frequency $f_\mathrm{r}$ at a readout power $P_\mathrm{in}=-44$~dBm measured with an integration time $\tau_\mathrm{int}=50~\mu$s and plotted together with the resonator response magnitude~$R$ in the even parity state. The SNR dependence shows the pronounced maximum at a readout frequency $f_\mathrm{r}=6.36206$~GHz, which is due to the signal term (the noise term is independent of $f_\mathrm{r}$). The signal term is determined by the change of the resonator response magnitude between the even and odd parity state ($\abs{R_\mathrm{e}-R_\mathrm{o}}$) resulting in the resonance-like shape, with a pronounced peak at $f_\mathrm{r}$ values, where the rate of change in~$R$ is strong.}
	\label{s8}
\end{figure}

\setcounter{myc}{9}
\begin{figure}[H]
	\includegraphics[width=0.5\columnwidth]{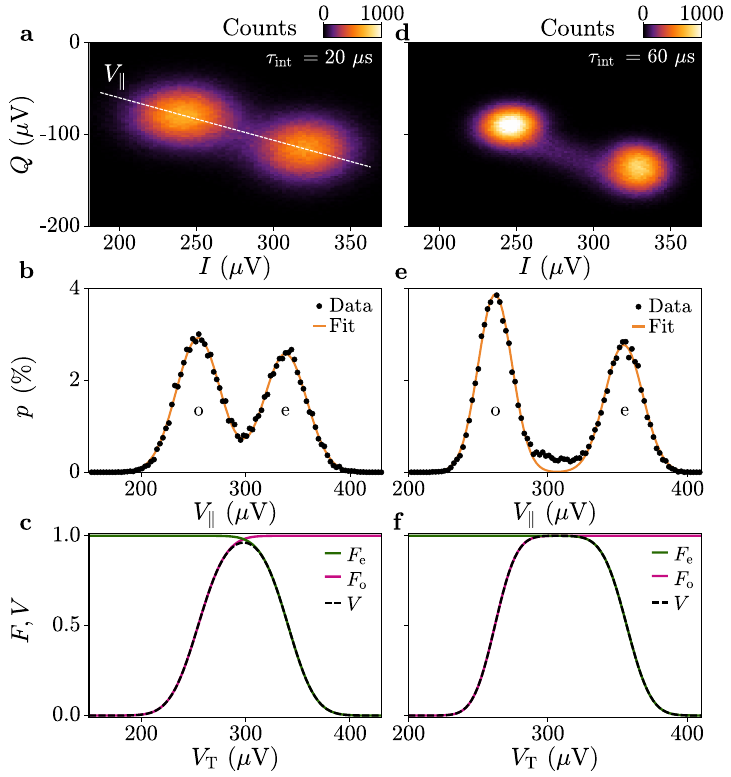}
	\centering
	\caption{\textbf{Readout fidelity determination.} \textbf{a}, Quadrature map of the two parity states for the integration time $\tau_\mathrm{int}=20~\mu$s plotted together with the line $V_\mathrm{\parallel}$ along the signal axis. \textbf{b}, Linecut from~\textbf{a} along $V_\mathrm{\parallel}$ normalized (black data points) and plotted together with the fit to the sum of two Gaussians (orange). \textbf{c}, Even and odd state fidelities $F_\mathrm{e},F_\mathrm{o}$ as well as visibility $V$ as a function of threshold voltage $V_\mathrm{T}$. The readout fidelity $F_\mathrm{e}$ ($F_\mathrm{o}$) is calculated by numerical integration of the fit in~\textbf{b} starting (stopping) at $V_\mathrm{T}$. The readout visibility is calculated as $V = F_\mathrm{e} + F_\mathrm{o} - 1$. \textbf{d}, Same as~\textbf{a} but for $\tau_\mathrm{int}=60~\mu$s. \textbf{e}, Same as~\textbf{b}, but with the linecut along the signal axis from \textbf{d}. The sum of two Gaussians cannot accurately fit the data due to a systematic error in the measurment. The slow sweep rate of the plunger gate voltage $V_\mathrm{P}$ during data acquisition resulted in a finite step width between even and odd parity state as seen in Fig.~\ref{fig3} in the Main Text. The finite step width can be mitigated by sweeping $V_\mathrm{P}$ faster, however, it does not affect the readout fidelity calculation presented here. \textbf{f}, Same as~\textbf{c} using the fit curve from~\textbf{e}.}
	\label{s9}
\end{figure}

\setcounter{myc}{10}
\begin{figure}[H]
	\includegraphics[width=0.5\columnwidth]{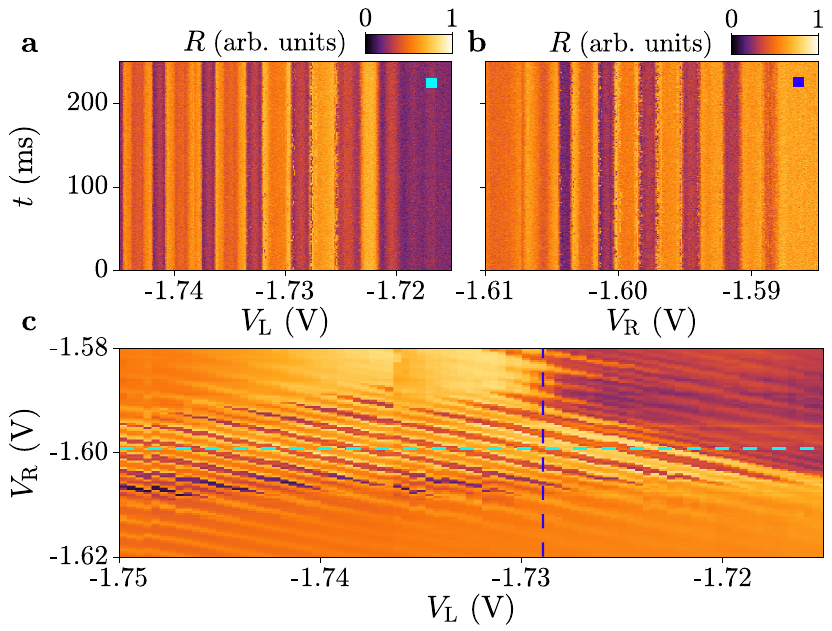}
	\centering
	\caption{\textbf{Gate-dependent parity switching measurements in Device~1.} \textbf{a}, Normalized resonator magnitude response $R$ as a function of time $t$ and left barrier gate voltage $V_\mathrm{L}$ at $V_\mathrm{R}=-1.599$~V, $V_\mathrm{P}=-1.6925$~V and $\it{\Phi}=\it{\Phi}_\mathrm{0}/\mathrm{2}$. At each $V_\mathrm{L}$ value, the parity was monitored for a time window of 900~ms ($\tau_\mathrm{int}=30~\mu$s, 30000 points). Parity switches are observed only close to transitions between even and odd ground state and only in a small gate voltage range around $V_\mathrm{L}=-1.73$~V. Deep in a fixed parity sector, the charging energy of the superconducting island suppresses the parity switches. \textbf{b}, Same as~\textbf{a}, but as a function of $V_\mathrm{R}$ at $V_\mathrm{L}=-1.729$~V. Here, the parity switches are observed at all transitions between the even and odd state. \textbf{c}, Overview of normalized resonator magnitude response $R$ as a function of barrier gate voltages $V_\mathrm{R}$ and $V_\mathrm{L}$. For $V_\mathrm{L}<-1.73$~V a transition from a 2$e$ periodic to an even-odd regime as a function of $V_\mathrm{R}$ is observed, as described in Extended Data Fig.~\ref{s5}. For more positive $V_\mathrm{L}$, in the even-odd regime the resonator magnitude response $R$ of the even parity state changes while $R$ of the odd parity state stays almost unaffected. The reason for this is unkown but parity switches detected in~\textbf{a} and~\textbf{b} appear only close to that gate voltage regime (turquoise and blue dashed lines show gate voltages used in~\textbf{a} and~\textbf{b}, respectively).}
	\label{s10}
\end{figure}

\newcounter{mycS}
\renewcommand{\thefigure}{S.\arabic{mycS}}

\subsection*{S1: Flux dependence and coupling factor $g$}
\begin{figure}[H]
	\includegraphics[width=0.5\columnwidth]{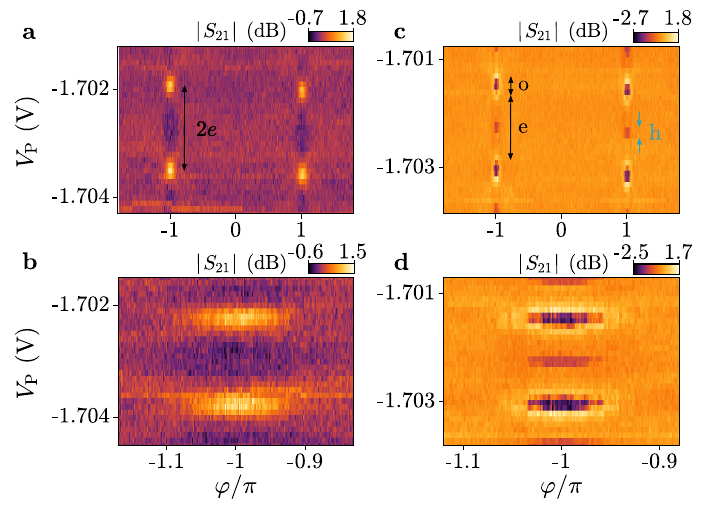}
	\centering
	\caption{\textbf{Flux dependence measurements of Device~1.} \textbf{a}, Resonator magnitude response $\lvert S_{21}\rvert$ as a function of total phase $\varphi$ across the superconducting leads and plunger gate voltage $V_\mathrm{P}$ in the regime with 2$e$ periodic Coulomb blockade oscillations ($V_\mathrm{L}=-1.755$~V, $V_\mathrm{R}=-1.617$~V). The Coulomb resonances are resolvable only around $\pi-$phase. \textbf{b}, Zoomed-in measurement of \textbf{a} around $\pi-$phase. \textbf{c}, Same as \textbf{a}, but in the even-odd regime ($V_\mathrm{L}=-1.755$~V, $V_\mathrm{R}=-1.609$~V). The odd parity state (o) as well as higher order transport effects (h) are observed only around $\pi-$phase. \textbf{d}, Zoomed-in measurement of \textbf{b} around $\pi-$phase.}
	\label{s11}
\end{figure}

Figures~\ref{s11}a,b shows the resonator response as a function of plunger gate voltage and magnetic flux in the $2e$-periodic regime. The even-odd regime, shown in Figs.~\ref{s11}c,d, is observed only when the energy of a subgap state is lower than the charging energy of the superconducting island ($\delta < E_\mathrm{C}$). In our system the subgap state is an Andreev bound state (ABS) with energy
\setcounter{mye}{1}
\begin{equation*}
	\delta(\varphi,\tau) = \pm \Delta \sqrt{1-\tau \sin^2 (\varphi / 2)},
\end{equation*}
where $\Delta$ is the induced superconducting gap of the Al, $\tau$ is the transmission of the ABS and $\varphi \approx 2 \mathrm{\pi} \it{\Phi}/ \it{\Phi}_\mathrm{0}$ is the phase difference across the two superconducting leads~\cite{BeenakkerABSFormula,BagwellABSFormula}. Since $\delta(\varphi,\tau)$ is minimal at $\varphi = \pi$ and $E_\mathrm{C}$ is independent of $\varphi$, the even-odd regime emerges at $\pi-$phase. Therefore, a flux dependence measurement allows to estimate $E_\mathrm{C}$ of the island if $\delta(\varphi,\tau)$ is known. We observe an anti-crossing between readout resonator and ABS in the resonator magnitude response $\lvert S_{21}\rvert$ which was measured as a function of $\varphi$ at similar gate voltages ($V_\mathrm{L}=-1.755$~V, $V_\mathrm{R}=-1.609$~V). From that measurement, we estimated $\tau \approx 0.98$ while $\Delta \approx 184~\mu$eV is known from two-tone spectroscopy measurements of similar devices~\cite{hinderling2023flip}. With these values, we estimated the charging energy of the superconducting island as $E_\mathrm{C} \approx 28~\mu$eV. 

\setcounter{mycS}{1}

\begin{figure}[H]
	\includegraphics[width=0.5\columnwidth]{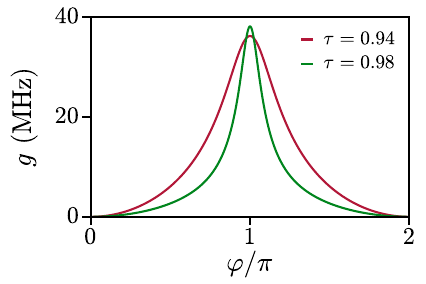}
	\centering
	\caption{\textbf{Coupling between readout resonator in Device~1.} Coupling factor $g$ between readout resonator and device loop as a function of total phase $\varphi$ across the superconducting leads for two transmission $\tau$ values of the bound state.}	
	\label{s12}
\end{figure}

The Coulomb resonance and blockade features in our measurements were observed only around $\pi-$phase, as shown in Fig.~\ref{s11}. This is due to the inductive coupling between the microwave resonator and the device loop, where circulating currents were mainly carried by a discrete Andreev bound state with transmission $\tau$. Thus, the coupling depended on $\varphi$ and $\tau$. Following Ref.~\cite{Janvier2015}, we calculate the coupling factor 
\begin{equation*}
	g(\varphi,\tau) = \frac{\sqrt{z}}{h} \frac{\Delta}{2} \frac{\delta(\pi,\tau)}{\delta(\varphi,\tau)} \tau \sin^2(\varphi/2)
\end{equation*}
with $h$ Planck's constant and $z=\pi M^2 \omega^2_\mathrm{R} / (Z_\mathrm{R} R_\mathrm{Q})$ a constant coupling parameter, where $M \approx 14.7$~pH is the calculated mutual inductance between resonator and device loop, $\omega_\mathrm{R} / (2 \pi) = 6.362$~GHz is the measured resonator frequency, $Z_\mathrm{R} \approx 56~\Omega$ is the characteristic resonator impedance and $R_\mathrm{Q} = h/(4e^2)$. The coupling $g$ peaks around $\pi-$phase and broadens with lower $\tau$ (see Fig.~\ref{s12}) consistent with the observations in Fig.~\ref{s11}. 

\setcounter{mycS}{2}

\subsection*{S2: Characterization of Device~2}
\begin{figure}[H]
	\includegraphics[width=0.5\columnwidth]{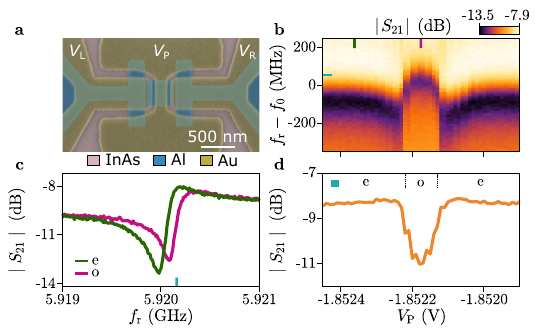}
	\centering
	\caption{\textbf{Parity detection measurements in Device~2.} \textbf{a}, False-colored scanning electron micrograph of Device~2. The gates (yellow) with voltages $V_\mathrm{P}$, $V_\mathrm{L}$ and $V_\mathrm{R}$ were used for tuning the chemical potential of the island and the coupling strength to the leads, respectively. \textbf{b}, Resonator response magnitude $\lvert S_{21} \rvert$ as a function of offset readout frequency $f_\mathrm{r}-f_\mathrm{0}$ and plunger gate voltage $V_\mathrm{P}$ at $V_\mathrm{R}=-1.0750$~V, $V_\mathrm{L}=-1.5031$~V and $\it{\Phi}=\it{\Phi}_\mathrm{0}/\mathrm{2}$. The modification of the resonator frequency indicates a transition between the even and odd parity states. \textbf{c}, Linecuts of resonator response at two marker positions indicated in~\textbf{b} showing the shift of the resonator between the even (e, green) and odd (o, pink) parity states. \textbf{d}, A linecut from~\textbf{b} at the turquoise marker position as a function of $V_\mathrm{P}$ emphasizes the change in the resonator magnitude response between even and odd parity state.}
	\label{s13}
\end{figure}

\begin{figure}[H]
	\includegraphics[width=0.5\columnwidth]{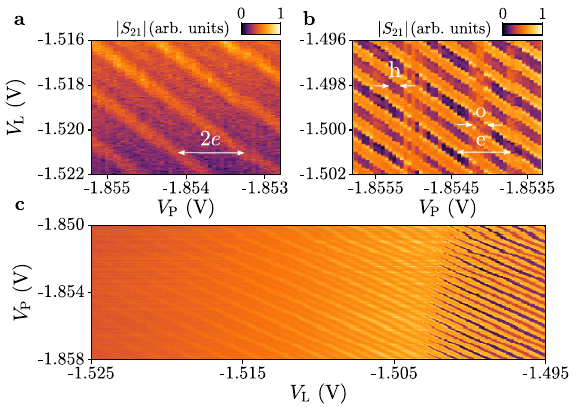}
	\centering
	\caption{\textbf{Charge state detection in Device~2.} \textbf{a}, Normalized resonator magnitude response $\lvert S_{21} \rvert$ as a function of plunger voltage $V_\mathrm{P}$ and left barrier voltage $V_\mathrm{L}$ at $\it{\Phi}=\it{\Phi}_\mathrm{0}/\mathrm{2}$ showing 2$e$ periodic Coulomb blockade oscillations as a function of both voltages. \textbf{a}, Same as~\textbf{a}, but showing parity effect with alternating even (e) and odd (o) parity sectors. Vertical stripes appeared due to flux drifts during the measurement. The coupling between resonator and device decreases rapidly as the flux drifts away from $\it{\Phi}=\it{\Phi}_\mathrm{0}/\mathrm{2}$ as discussed in Fig.~\ref{s12}. Higher order transport features (h)~\cite{PhysRevLett.118.137701} are more pronounced compared to Device~1, because the coupling to the leads was larger in Device~2. For an island with small charging energy $E_\mathrm{C}$ the bound state energy $\delta$ must be small to enter the even-odd regime leading to a strong coupling to the leads. \textbf{c}, Same as~\textbf{a} and~\textbf{b}, but sweeping $V_\mathrm{L}$ in a wide range. We observe (i) pure 2$e$ periodic Coulomb oscillations ($V_\mathrm{L}<-1.515$~V), (ii) 2$e$ periodic Coulomb oscillations with higher order transport processes (for $V_\mathrm{L}$ between $-1.515$~V and $-1.503$~V) and (iii) even-odd regime ($V_\mathrm{L}>-1.503$~V).}
	\label{s14}
\end{figure}

\begin{figure}[H]
	\includegraphics[width=0.5\columnwidth]{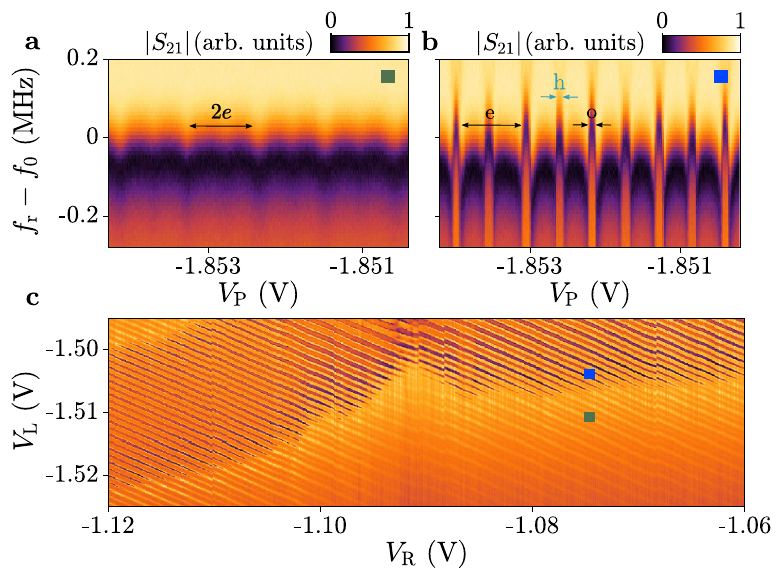}
	\centering
	\caption{\textbf{Additional measurement on Device~2.} \textbf{a}, Normalized resonator magnitude response $\lvert S_{21} \rvert$ as a function of offset readout frequency $f_\mathrm{r}-f_\mathrm{0}$ and plunger gate voltage $V_\mathrm{P}$ at $V_\mathrm{R}=-1.075$~V, $V_\mathrm{L}=-1.514$~V and $\it{\Phi}=\it{\Phi}_\mathrm{0}/\mathrm{2}$. The 2$e$ periodic Coulomb blockade oscillations are consistent with the charging energy of the island being below the bound state energy ($E_\mathrm{C}<\delta$). At Coulomb resonances, a shift in the resonator frequency is observed due to enhanced circulating current in the device loop, which changes the inductance of the device. \textbf{b}, Same as \textbf{a}, but at  $V_\mathrm{L}=-1.506$~V. Here, $\delta < E_\mathrm{C}$ that allows the number of quasiparticles on the superconducting island to be even or odd. The inductance of the even~(e) and odd~(o) parity states are different resulting in distinct resonator responses. The modification of the even state is more pronounced than in Fig.~\ref{s5} because a smaller coupling to the leads suppresses the higher order transport processes. \textbf{c}, Normalized resonator magnitude response $\lvert S_{21} \rvert$ as a function of barrier gate voltages $V_\mathrm{L}$ and $V_\mathrm{R}$ at $V_\mathrm{P}=-1.855$~V. The gate-voltage map shows Coulomb blockade oscillations because both barrier gates are able to tune the chemical potential of the superconducting island. We observe two isolated bound states, one on the left side and one in the upper right area of the gate-voltage map, which were relevant and tunable with $V_\mathrm{L}$ and $V_\mathrm{R}$, respectively. When at the same gate voltages the energy of both bound states is lower than $E_\mathrm{C}$, i.e. $V_\mathrm{L}=-1.495$~V and $V_\mathrm{R}=-1.095$~V, the even-odd regime becomes 1$e$ periodic.}
	\label{s15}
\end{figure}

A second device, called Device~2, was investigated for this study. Device~2 was identical to Device~1 except for the size of the superconducting island, which was approximately 400~nm by 400~nm. Device~2 was fabricated on the same chip as Device~1, but was coupled to a different resonator with resonance frequency 5.920~GHz. A scanning electron micrograph of Device~2 is shown in Fig.~\ref{s13}a. Basic parity readout is demonstrated in Figs.~\ref{s13}b-d. An extended gate-dependent study of Device~2 is shown in Figs.~\ref{s14} and ~\ref{s15}, with results similar to those obtained on Device~1.

\setcounter{mycS}{3}

\setcounter{mycS}{4}

\setcounter{mycS}{5}

\end{document}